\documentclass[12pt,preprint2]{aastex}

\usepackage[margin=1.0 in]{geometry}
\usepackage{times}
\usepackage{amsmath}
\usepackage{rotating,tabularx}

\newcommand{\bc}{\begin{center}}
\newcommand{\ec}{\end{center}}
\newcommand{\be}{\begin{equation}}
\newcommand{\ee}{\end{equation}}
\newcommand{\bfig}{\begin{figure}}
\newcommand{\efig}{\end{figure}}
\newcommand{\m}{\mbox}

\received{?}
\revised{?}
\accepted{?}
	
\usepackage{amssymb}

\begin{document}
\title{UV Surface Environments and Atmospheres of Earth-like Planets Orbiting White Dwarfs}
\author{Thea Kozakis, Lisa Kaltenegger}
\affil{Carl Sagan Institute, Cornell University, Ithaca, New York, USA}
\and
\author{D.\ W.\ Hoard}
\affil{Eureka Scientific, Oakland, California, USA}
\date{}

\begin{abstract}
An Earth-like exoplanet orbiting a white dwarf  would be exposed to different UV environments than Earth, influencing both its atmospheric photochemistry and UV surface environment. Using a coupled 1D climate-photochemistry code we model atmospheres of Earth-like planets in the habitable zone of white dwarfs for surface temperatures between 6000~K and 4000~K, corresponding to about 7~billion years of white dwarf evolution, as well as discuss the evolution of planetary models in the habitable zone during that evolution.

\end{abstract}

\maketitle

\section{Introduction}

Exoplanets have not yet been discovered orbiting white dwarfs, but have been found around pulsars, indicating that it is possible to have planetary bodies orbiting stellar remnants \citep{wols92}.  Exoplanet searches are underway around white dwarfs (WDs) (e.g.\ \cite{fult14,foss15,vera15,xu15}), as the similarity with Earth's size should make Earth-sized exoplanets transiting WDs easier to detect and characterize than such planets around much larger main sequence stars (e.g.\ via atmospheric profile measurements from transit observations, ``weather" modeling from orbital light curves, direct measurement of atmospheric constituents with spectroscopy, etc.) \citep{agol11}. Studies of close-by WDs with NASA's K2 mission \citep{howe14} constrain the occurrence of Earth-sized habitable zone (HZ) planets in those systems to be $<$ 28\% \citep{slui17}.

Assuming that an Earth-like planet could form or survive around a WD, the WD cooling process will provide a changing luminosity as well as UV environment, which affects an orbiting planet's temperature, atmospheric photochemistry, and UV surface flux. The luminosity of a cool WD evolves slower than in its initial hot phase, thus cool WDs could provide stable environments for potentially habitable planets \citep{agol11}. Several studies have suggested that the unique UV environment would be high enough to sustain complex chemical processes necessary for Earth-like life, while not being strong enough to damage DNA \citep{mccr71,foss12} using estimates assuming present day Earth atmospheres.

Multiple teams have shown that a fraction of WDs show evidence of recent heavy metal pollution, which could signal the existence of either disks or planets (e.g.\ \cite{koes06,koes14,hame16,klei11,mala16}). For reviews of WD debris disks and pollution see \cite{jura14} and \cite{fari16}, and for potential post-main sequence planetary evolution see \cite{vera16}. 

\begin{figure*}[h!]
\includegraphics[scale=0.4]{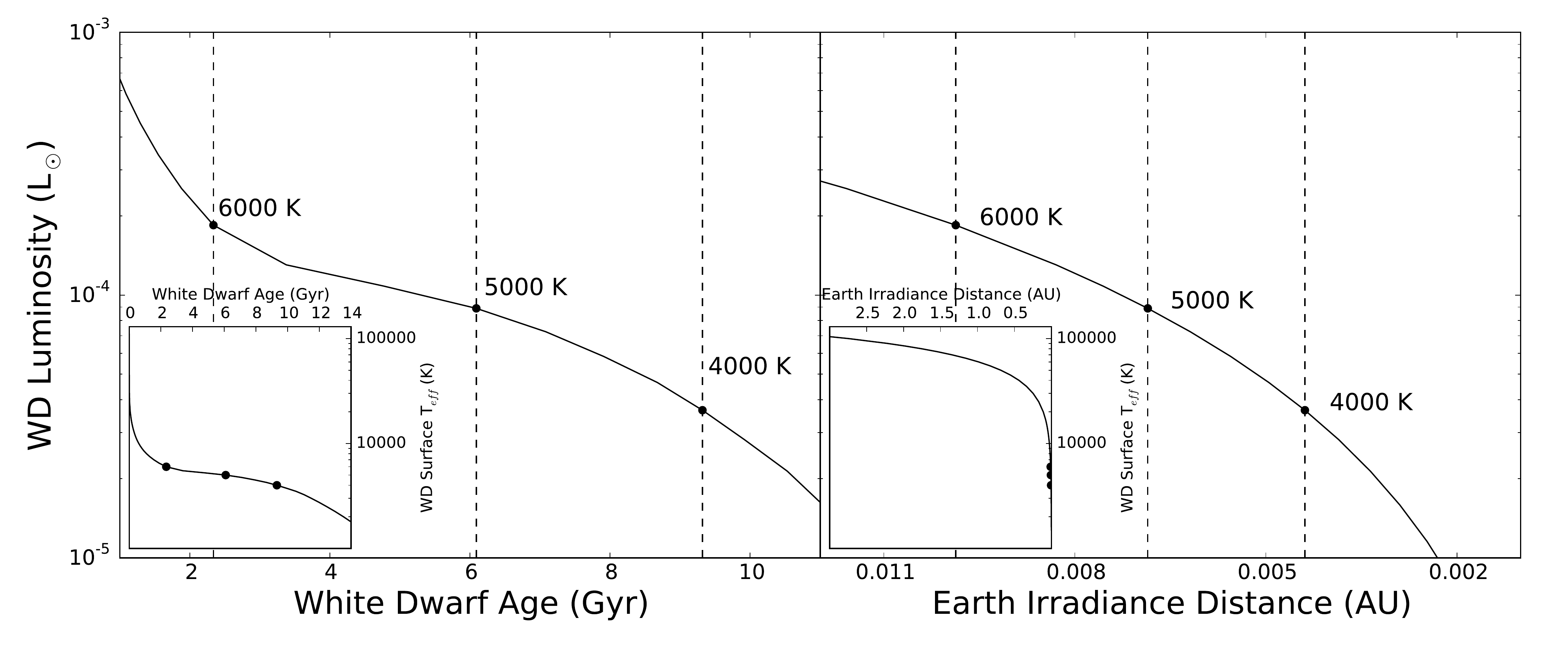}
\caption{White dwarf (WD) luminosity evolution calculated using cooling models from \cite{berg01} for a 0.6~M$_\odot$ WD, showing (left) WD luminosity (L$_{WD}$) versus age of a 0.6~M$_\odot$ WD model and (right) the 1 AU equivalent orbital distance, where a planet would receive the same irradiance as Earth, for the same WD model. The three points (6000~K, 5000~K, and 4000~K) indicate the WD evolution for the planet models discussed. The inset panels show the same quantities plotted over a larger parameter ranges using the WD's surface temperature on the y-axis instead of the luminosity (see text).
\label{WD_cooling}}
\end{figure*}

\begin{figure*}[h!]
\includegraphics[scale=0.45]{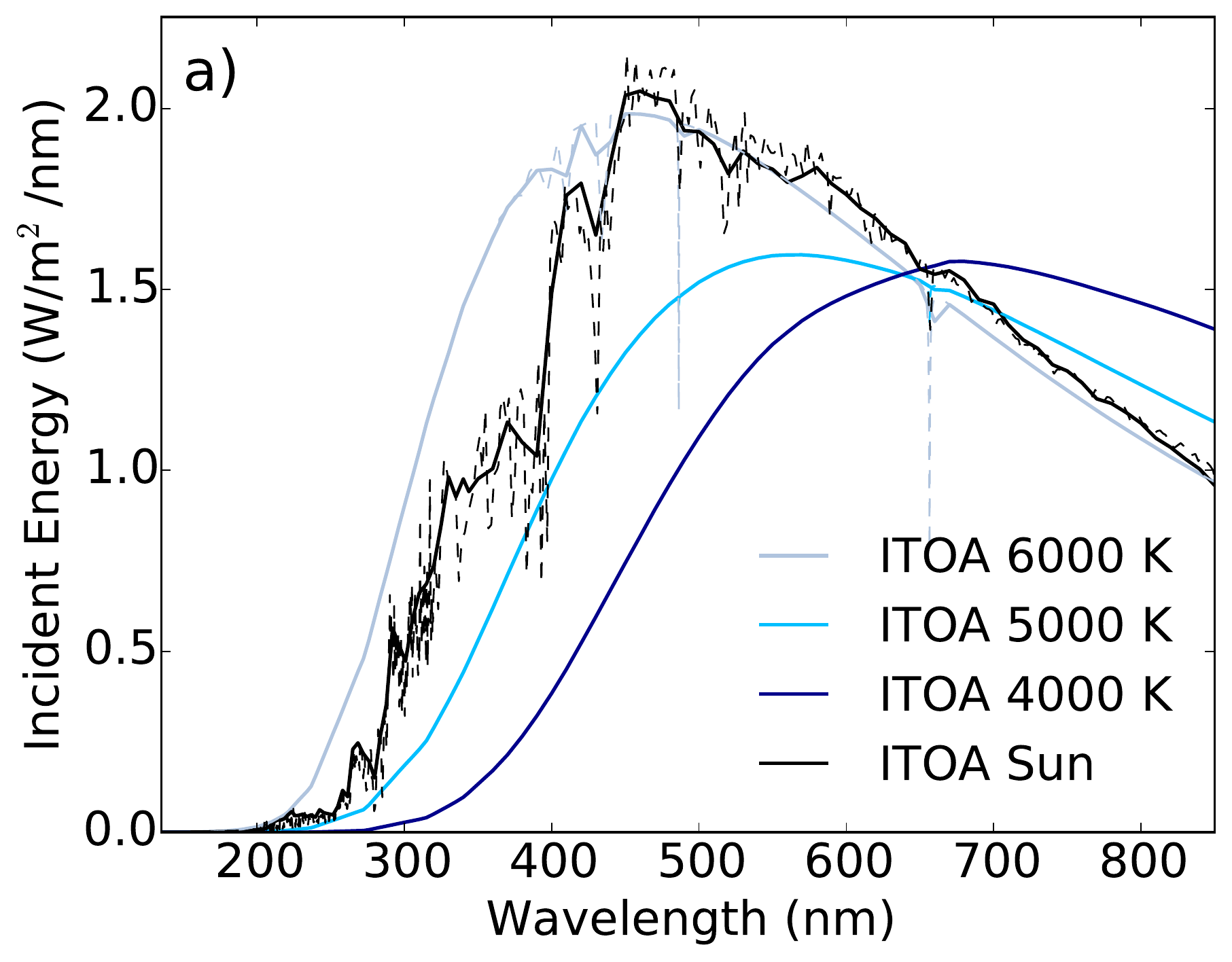}
\includegraphics[scale=0.45]{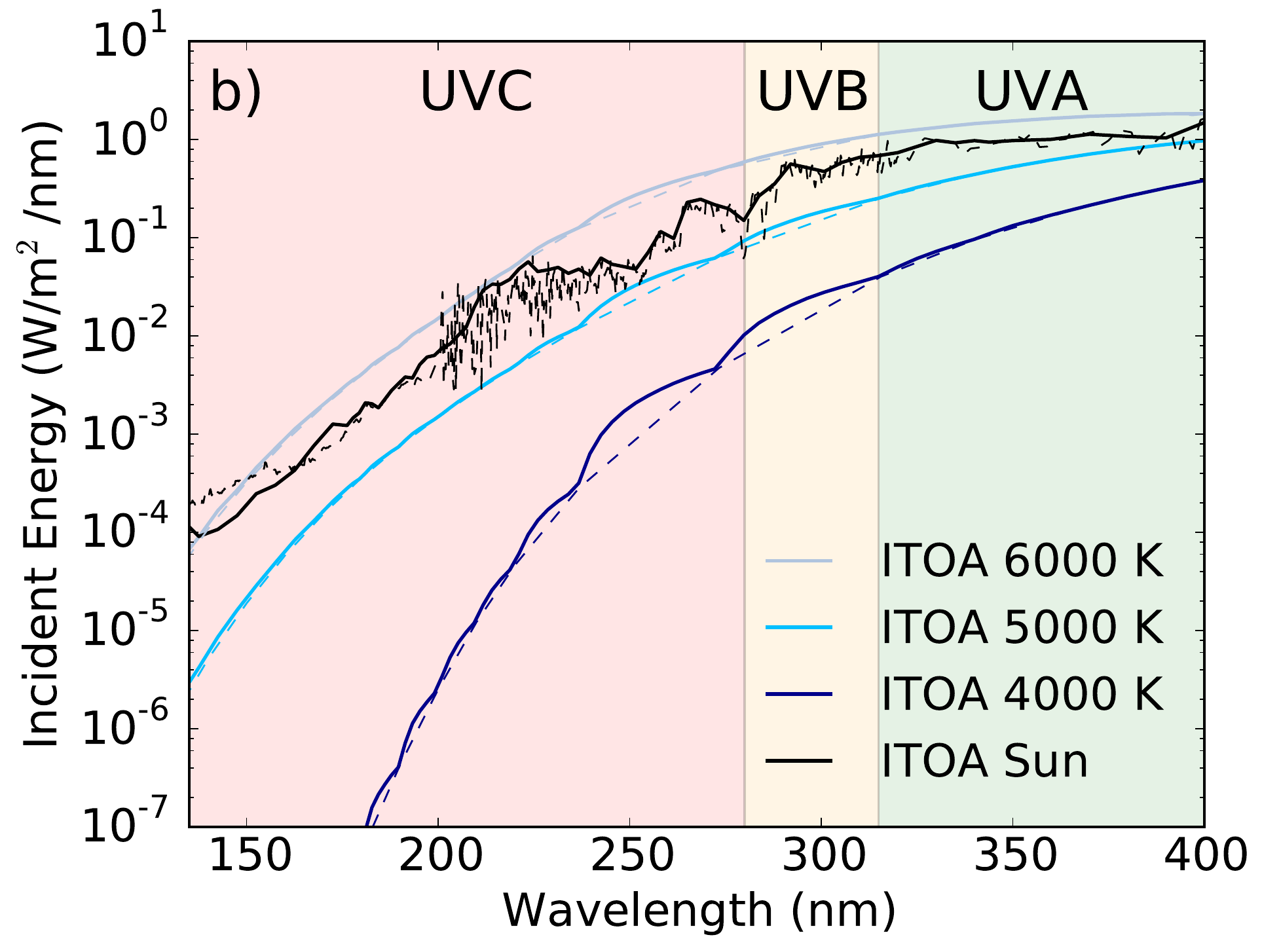}
\caption{Comparisons of the irradiance at the top of the atmosphere (ITOA) for planets orbiting a WD compared to the Earth-Sun system both at high resolution (dashed lines) and binned to the wavelength grid of the climate and photochemistry model resolution (solid lines).  All spectra are scaled to the level of present day Earth's irradiance. \label{WD_models}}
\end{figure*}

Transit simulations using an unchanged present-day Earth-analog atmosphere composition suggest that the depth of Earth's biosignatures around a WD would be very strong and detectable by future missions such as \emph{JWST} \citep{loeb13}. However a WD's spectral energy distribution (SED) can be significantly different from our present-day Sun. To model the effects of the WD's emitted flux on the atmosphere as well as the surface UV environment of an exoplanet, we use incident WD SED models at different points of a 0.6~M$_\odot$WD's evolution (Figure~\ref{WD_cooling}) in our models between 6000~K and 4000~K (Figure~\ref{WD_models}).

We model planets orbiting WDs at three points in their evolution (Figure~\ref{WD_models}), with surface temperatures at 6000~K, 5000~K and 4000~K, during which time the WD's luminosity  does not change significantly compared to the early luminosity change in a WD's evolution (see Figure~\ref{WD_cooling}). That surface temperature change corresponds to about 7~billion years (Gyr) for a 0.6~M$_\odot$WD. We consider planets with eroded atmospheres as well as with higher surface pressure (e.g.\ super-Earths). We show the atmospheric structure and chemical composition as well as the UV surface fluxes at biologically relevant wavelengths.

\section{Methods \label{methods}}

\subsection{Planet model: \emph{EXO-Prime}}

We use \emph{EXO-Prime} (see e.g.\ \cite{kalt10}); a coupled 1D radiative-convective atmosphere code developed for rocky exoplanets. The code is based on iterations of a 1D~climate model \citep{kast86,pavl00,haqq08}, a 1D~photochemistry model \citep{pavl02,segu05,segu07}, which are run to convergence (see details in \cite{segu05}). \emph{EXO-Prime} models exoplanet atmospheres and environments depending on the stellar and planetary conditions, including the UV radiation that reaches the surface and the planet's reflection, emission and transmission spectrum. We divide the atmosphere in 100~plane parallel layers for our model up to an altitude of 60~km (or a pressure of 1~mbar) using a stellar zenith angle of 60~degrees.

Visible and near-IR shortwave fluxes are calculated with a two stream approximation including atmospheric gas scattering \citep{toon89}, and longwave fluxes in the IR region are calculated with a rapid radiative transfer model (RRTM). A reverse-Euler method within the photochemistry code (originally developed by \cite{kast85}) contains 220 reactions to solve for 55 chemical species. 

First, we scale the incident WD flux at the top of the model planetary atmospheres to the total integrated flux Earth receives from the Sun (S$_{eff}$) to model how the WD's irradiance changes the planetary environment compared to Earth. We then bin the high resolution spectra to the resolution of the wavelength grid of the climate/photochemistry models (both shown in Figure~\ref{WD_models}). We model planets with surface pressures of 0.3~bar (e.g.\ eroded atmosphere), 1~bar (Earth analogue), and 1.5~bar and 2~bar.

In our planetary models we keep the planetary outgassing rates constant for H$_2$, CH$_4$, CO, N$_2$O, and CH$_3$Cl and maintain the mixing ratios of O$_2$ at 0.21 and CO$_2$ at 3.55$\times10^{-6}$ to be able to compare the effect of the irradiation on the planetary atmosphere, with a varying N$_2$ concentration that is used as a fill gas to reach the set surface pressure of the model (following \cite{segu03,segu05,rugh13,rugh15,rugh15b,rugh18}). Note that by keeping the outgassing rates constant, lower surface pressure atmosphere models initially have slightly higher mixing ratios of chemicals with constant outgassing ratios than higher surface pressure models.

\subsection{White Dwarf Model Spectra}

WDs are unique stellar environments with very high surface gravities and extremely dense interiors and physical conditions. They have no internal heat source, and thus cool off over time (Figure~\ref{WD_cooling}). We use WD spectral energy distribution models (SED) calculated as described in \cite{saum14} (Figure~\ref{WD_models}) as irradiation for the planetary models (see e.g. \cite{berg97,kowa06,kili09a,kili09b,giam12,saum14} for detailed discussion on WD spectra). The WD models assume a pure H atmospheric composition with a surface gravity of  $log~g~=~8.0$ in creating the models for the average mass of WD in the field (0.6~M$_\odot$; e.g.\ \cite{kpk2016}).  The WD radius is then recovered by finding the intersection of the standard surface gravity formula with the WD mass-radius-(temperature) relation for C-O core WDs with thick H layers discussed by \cite{pgm2017}(see their Figure 9; also see \cite{ba1999,fbb2001,pgm2017} extrapolated to T$<$10,000 K. The radius of the WD in our model is 0.0128$\pm$0.0001 R$_\odot$ for surface temperatures of 6000~K to 4000~K.  WD spectra are similar to black bodies with only Balmer absorption lines for surface temperatures greater than 5000~K, where hydrogen becomes neutral as shown in Figure~\ref{WD_models}. We use models from \cite{berg01} for the temperature evolution of the WD (see Figure~\ref{WD_cooling}).

\subsection{Photochemistry of some biologically interesting species}
Some atmospheric species exhibit noticeable  features in our planet's spectrum as a result directly or indirectly from biological activity. The main ones are oxygen (O$_2$), ozone (O$_3$), methane (CH$_4$), nitrous oxide (N$_2$O) and methyl chloride (CH$_3$Cl) (see e.g.\ \cite{dema12,kalt17}). We summarize the most important reactions that influence these species in Earth's atmosphere here. Many reactions in Earth's atmosphere are driven by the Sun's UV flux. Ozone and O$_2$ are created with UV photons through the Chapman reactions \citep{chap30},
\begin{equation}
\begin{aligned}
\m{O}_2 + \m{h}\nu \rightarrow \m{O + O } (\lambda < 240\ \m{nm}), \\
\m{O + O}_2 + M \rightarrow \m{O}_3 + M, \\
\m{O}_3 + \m{h}\nu \rightarrow \m{O}_2 \m{ + O}\ (\lambda < 320\ \m{nm}),\\
\m{O}_3 + \m{O} \rightarrow \m{2O}_2, 
\end{aligned}
\label{O3}
\end{equation}
where $M$ is a background molecule such as N$_2$.  These reactions are primarily responsible for ozone production on present day Earth.  A higher UV flux additionally increases the primary source reactions of tropospheric hydroxyl (OH) for 300~$<\lambda<$~320~nm in the troposphere,
\begin{equation}
\begin{aligned}
\mbox{O}_3 + \mbox{h}\nu \rightarrow \mbox{O(}^1\mbox{D) + O}_2, \\
\mbox{O(}^1\mbox{D) + H}_2\mbox{O} \rightarrow \mbox{OH + OH}, 
\end{aligned}
\label{OH}
\end{equation}
reducing O$_3$ and H$_2$O. Increased OH is the primary sink for H$_2$, CH$_4$, CH3Cl, and CO abundances.  

Methane (CH$_4$) is a reducing gas that has a lifetime of 10-12 years in present day Earth's atmosphere \citep{houg04} due to reactions with oxidizing species.  It has both natural (termites, wetlands) and anthropogenic sources (rice agriculture, natural gas). It is oxidized via 
\begin{equation}
\label{CH4_ox}
\m{CH}_4 + 2\m{O}_2 \rightarrow  \m{CO}_2 + 2\m{H}_2\m{O},
\end{equation}
creating H$_2$O and H$_2$ via photolysis,
\begin{equation}
\label{CH4_ph}
\m{CH}_4 + \m{h}\nu  \rightarrow  \m{CH}_2 + \m{H}_2.
\end{equation}
Its main sink is due to the reaction,
\begin{equation} 
\label{CH4}
\mbox{OH} + \mbox{CH}_4 \ \rightarrow \ \mbox{CH}_3 + \mbox{H}_2\mbox{O},
\end{equation}
in the troposphere, and photolysis in the stratosphere. This reaction also creates H$_2$O in the stratosphere, above where it is shielded by the ozone layer.  

On Earth both nitrous oxide (N$_2$O) and methyl chloride (CH$_3$Cl) are primarily produced by life, and are depleted by higher amounts of UV.  N$_2$O is a greenhouse gas that is extremely effective when trapping heat.  It is produced naturally in soil through nitrification and denitrification, and its anthropogenic source is from agriculture.  Photolysis of N$_2$O occurs for $\lambda <$  220 nm, and is its main sink on present day Earth.  Increased O$_3$, e.g.\ due to UV irradition, is also a sink of N$_2$O,
\begin{equation}
\label{N2O}
\m{N}_2\m{O + O(}^1\m{D)} \rightarrow \m{2NO},
\end{equation}
creating NO. It depletes ozone via
 \begin{equation}
\m{NO + O}_3 \rightarrow \m{NO}_2 + \m{O}_2.
\end{equation}
On Earth CH$_3$Cl is naturally produced in oceans via light interacting with sea foam chlorine and biomass and in small amounts by phytoplankton. CH$_3$Cl reacts with OH creating Cl, a component of chlorofluorocarbons which damage the ozone layer.  It is converted into chlorine via the reactions,
\begin{equation}
\begin{aligned}
\m{CH}_3\m{Cl + OH} \rightarrow \m{Cl + H}_2\m{O}, \\
\m{CHCl + h}\nu \rightarrow \m{CH +Cl}, \\
\m{CH}_3\m{Cl + Cl} \rightarrow \mbox{HCl + Cl}.
\end{aligned}
\label{CH3Cl}
\end{equation}

\section{Results \label{results}}

\subsection{Time evolution of a white dwarf's habitable zone}
During the cooling process of a WD, its surface temperature as well as its overall luminosity decreases. This, in turn, influences the orbital distance of its HZ. The HZ is the circular region around one or multiple stars in which standing bodies of liquid water could exist on a rocky planet's surface (e.g.\ \cite{kast93,kalt13,hagh13,kopp13,kopp14,rami17}) and facilitate the detection of possible atmospheric biosignatures (see e.g.\ \cite{kalt17}). The classical, conservative N$_2$-CO$_2$-H$_2$O HZ is defined by the greenhouse effect of two gases: CO$_2$ and H$_2$O vapor. The inner edge corresponds to the distance where mean surface temperatures exceed the critical point of water ($\sim$647~K, 220~bar), triggering a runaway greenhouse that leads to rapid water loss to space on very short timescales (see \cite{kast93} for details). Toward the outer edge of the classical HZ weathering rates decrease, allowing atmospheric CO$_2$ concentrations to increase as stellar insolation decreases. At the outer edge, condensation and scattering by CO$_2$ outstrips its greenhouse capacity, the so-called maximum greenhouse limit of CO$_2$.

\begin{figure*}[h!]
\includegraphics[scale=0.42]{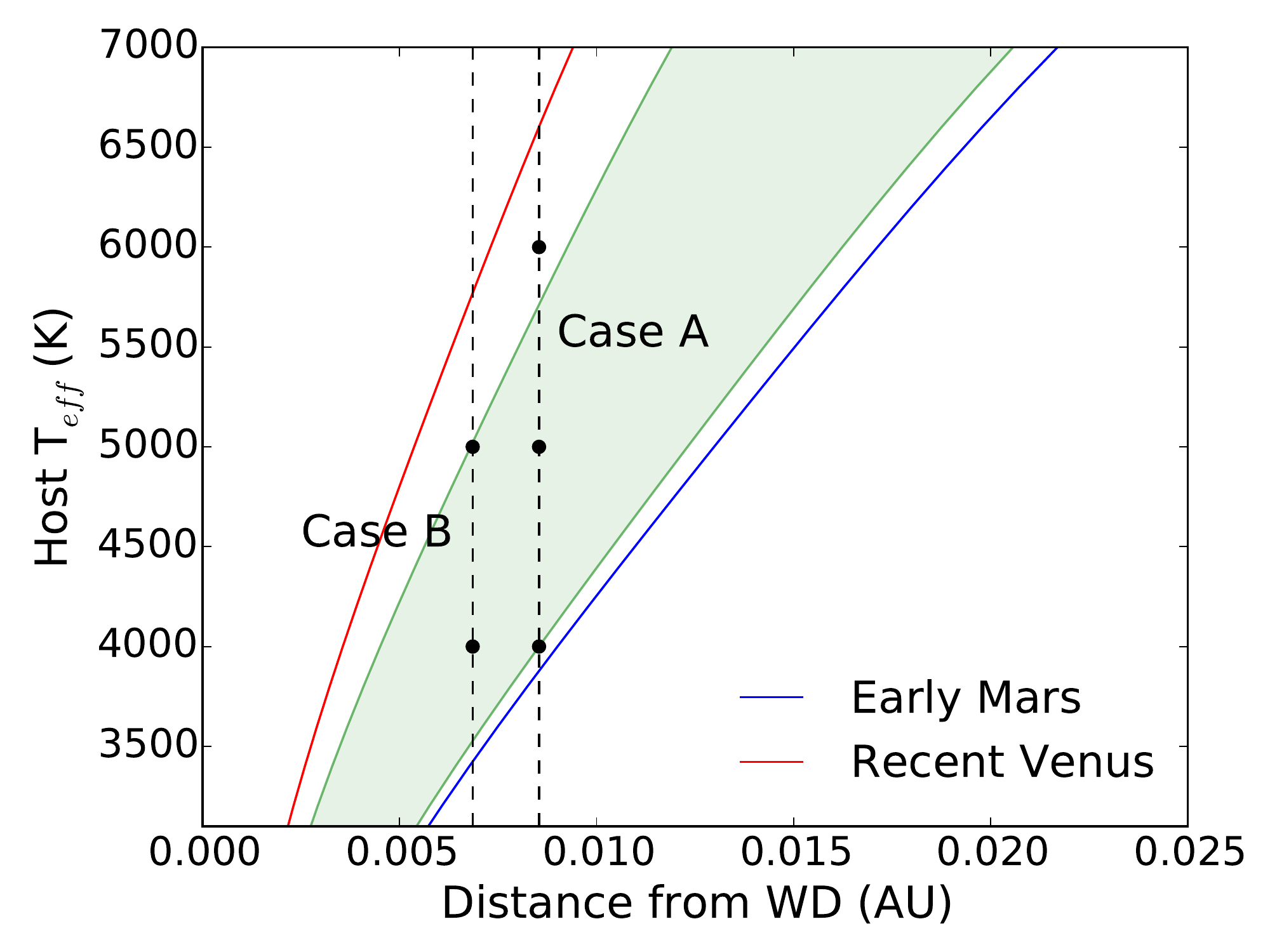}
\label{WD_HZ_au}
\includegraphics[scale=0.42]{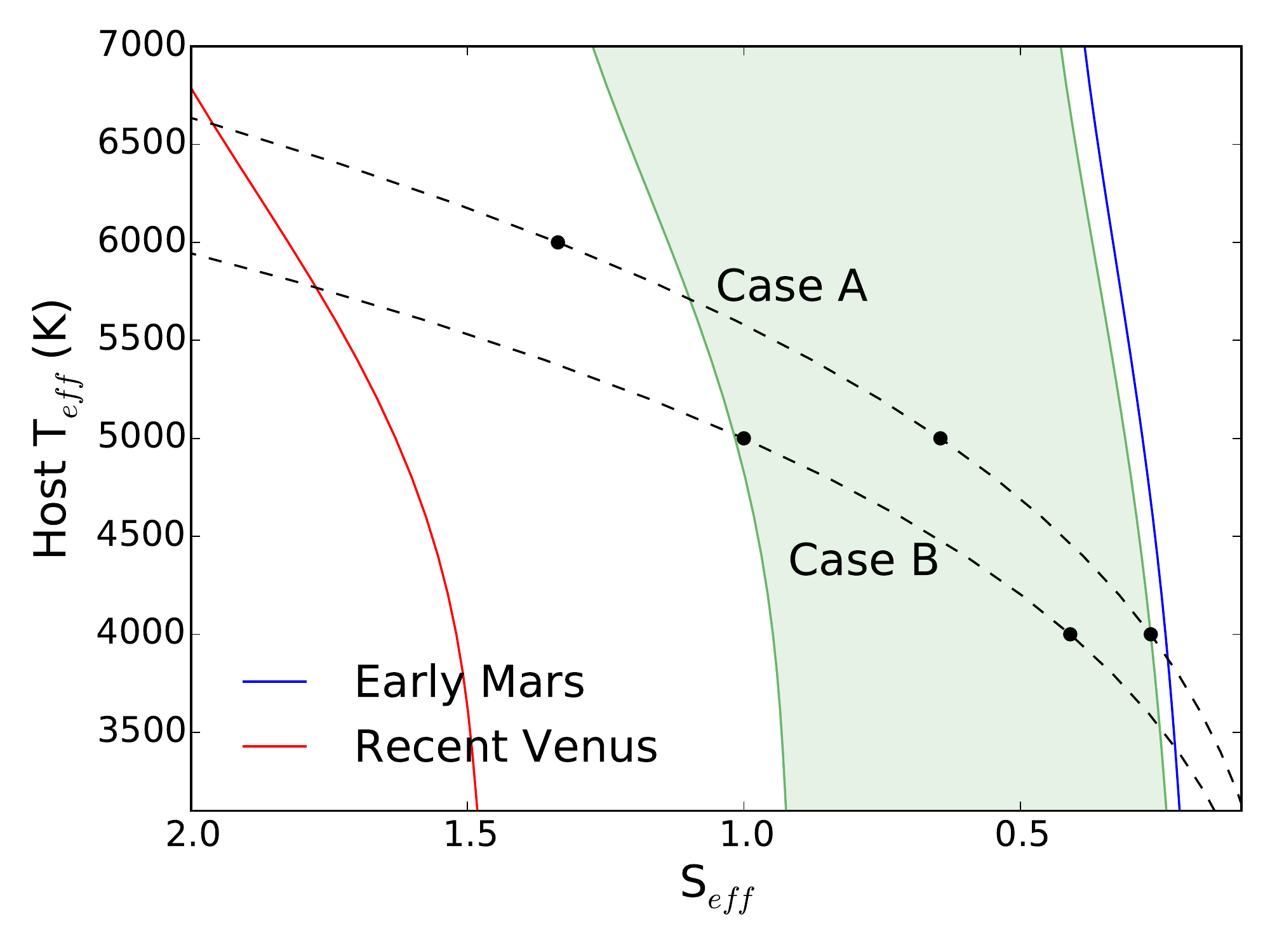}
\label{WD_HZ_seff}
\caption{Size of a WD's HZ (left) as well as the corresponding range of WD irradiance, normalized to the value for Earth (S$_{eff}$) (right). The shaded green region represents the classical HZ for our WD model (following \cite{kopp14}). The red (recent Venus) and blue (early Mars) lines show the limits of the empirical HZ (following \cite{rami17}). Two case studies for a planet in the WD's HZ, Case~A and~B, are indicated in both panels. The black dots indicate the specific WD surface temperatures and planetary distances/S$_{eff}$\  values we model in this study.
\label{WD_HZ}}
\end{figure*}

The cooling of a WD translates into an inward shift of the orbital distances of the WD HZ, (Figure~\ref{WD_HZ}a). In addition to the decrease in overall irradiance (S$_{eff}$) as the WD cools, some of the change in the HZ orbital distance is also caused by the shift of the peak wavelength of emission of the WD SED to redder wavelengths, which heat the surface of the planet more efficiently than bluer light (e.g.\ \cite{kast93}).  A 0.6~M$_\odot$WD spends about 7~Gyr cooling from 6000~K to 4000~K, providing a phase during which its luminosity does not rapidly change and thus it could provide temperate conditions for an orbiting planet. The size of the WD HZ is shown in Figure~\ref{WD_HZ}a as well as the corresponding irradiance of a planet, normalized to the value for Earth (S$_{eff}$) shown in Figure~\ref{WD_HZ}b. Because of the small size of a WD compared to the Sun, the WD HZ is a factor of $\sim$100 to~$\sim$1000 times closer to the WD than Earth is to the Sun.

An alternative HZ limit that is not based on atmospheric models (like the classical HZ) but on empirical observations of our Solar System is also shown in Figure~\ref{WD_HZ} for comparison. The inner edge of the empirical HZ is defined by the stellar flux received by Venus when we can exclude the possibility that it had standing water on the surface (about 1~Gyr ago), equivalent to a stellar flux of S$_{eff}$~=~1.77, corresponding to a distance of 0.75~AU for Earth's current Solar flux \citep{kast93}. The ``early Mars" limit is based on observations suggesting that the Martian surface may have supported standing bodies of water ~$\sim$3.8~Gyr ago, when the Sun was only 75\% as bright as today. For our solar system, S$_{eff}$~=~0.32 for this limit, corresponding to a distance of $
\sim$1.77 AU (e.g.\ \cite{kast93}).

We first explore the range of conditions for Earth-like planets orbiting WDs of different temperatures seen at one point in time (see Sections~3.2 and~3.3). Then we model two case studies, A and B, as shown in Figure~\ref{WD_HZ}, which explore the environment of a planet in the HZ of a WD as it cools (see Section~3.4).

\subsection{Planetary photochemistry environments for different stages in a WD's evolution}
The lack of chromospheric activity for a WD at different stages in its evolution causes photochemical differences in the atmospheres of planets orbiting it compared to Earth. We modeled planets with different surface pressures, which receive an equivalent total irradiance as Earth from the Sun (S$_{eff}$) from a WD at three different stages in its evolution: for WD surface temperatures of 6000~K, 5000~K and 4000~K.  We model planets with surface pressures ranging from 2~bar to 0.3~bar. Table~\ref{model_summary} summarizes the model planet surface temperature and integrated overall ozone column depth (ozone column depth) for the different planetary models as well as our Earth model for comparison. Figure~\ref{all_photochemistry} shows the changes in temperature as well as the mixing ratio for O$_3$, CH$_4$, H$_2$O, CH$_3$Cl and N$_2$O for the different models, compared to Earth.

\begin{table*}[t!]
\begin{center}
\caption{Model summary for Earth-equivalent irradiance \label{model_summary}}
\begin{tabular}{cccc}
Host T$_{\footnotesize \mbox{eff}}$ & Pressure &  Surface  & Ozone Column  \\
(K) & (bar) &T$_{\footnotesize \mbox{eff}}$ (K) & Depth (cm$^{-2}$) \\
\hline
Present day Earth & 1.0 & 288.2 & 5.4$\times10^{18}$\\
\hline
6000 & 0.3 &		280.7 &		3.6$\times10^{18}$\\
6000 & 1.0 & 		285.6 &		5.7$\times10^{18}$\\
6000 & 1.5 &		288.0 &		6.5$\times10^{18}$\\
6000 & 2.0 &		289.0 &		6.9$\times10^{18}$\\
\hline
5000 & 0.3 &		282.4 &		2.6$\times10^{18}$\\
5000	 & 1.0 &		290.8 &		3.9$\times10^{18}$\\
5000 & 1.5 &		295.0 &		4.2$\times10^{18}$\\
5000 & 2.0 &		298.1 &		4.5$\times10^{18}$\\
\hline
4000 & 0.3 &		282.8 &		1.1$\times10^{18}$\\	
4000 & 1.0 &		294.3 &		1.8$\times10^{18}$\\
4000 & 1.5 &		300.4 &		2.1$\times10^{18}$\\
4000 & 2.0 &		305.1 &		2.3$\times10^{18}$\\	
\hline
\label{planets}
\end{tabular}
\end{center}
\end{table*}

Figure~\ref{all_photochemistry} shows that in the model atmospheres H$_2$O photolysis increases with higher UV levels in the upper atmosphere, where H$_2$O is not shielded from incoming photons below the ozone layer. N$_2$O is depleted by photolysis, with a decreasing mixing ratio toward the top of the atmosphere, but remains well mixed beneath the ozone layer, where it is shielded. All higher surface pressure models show increased amounts of CH$_4$, N$_2$O, and CH$_3$Cl in the upper atmosphere. Atmospheric H$_2$O increases with higher surface temperatures. 

{The WD cooling process provides a changing luminosity as well as UV environment at a set orbital distance, as seen in Figure~\ref{WD_cooling}. The decrease in UV incident flux leads to an overall decrease in the ozone level. Compared to present day Earth's integrated overall ozone column depth, a higher surface pressure atmosphere receiving similar UV irradiance has a higher ozone column depth, as seen in the values for our planet models with surface pressures above 1~bar orbiting a WD with a surface temperature of 6000~K. Table~\ref{model_summary} shows the absolute values for our models as well as present day Earth's.

\begin{figure*}[h!]
\includegraphics[scale=0.85]{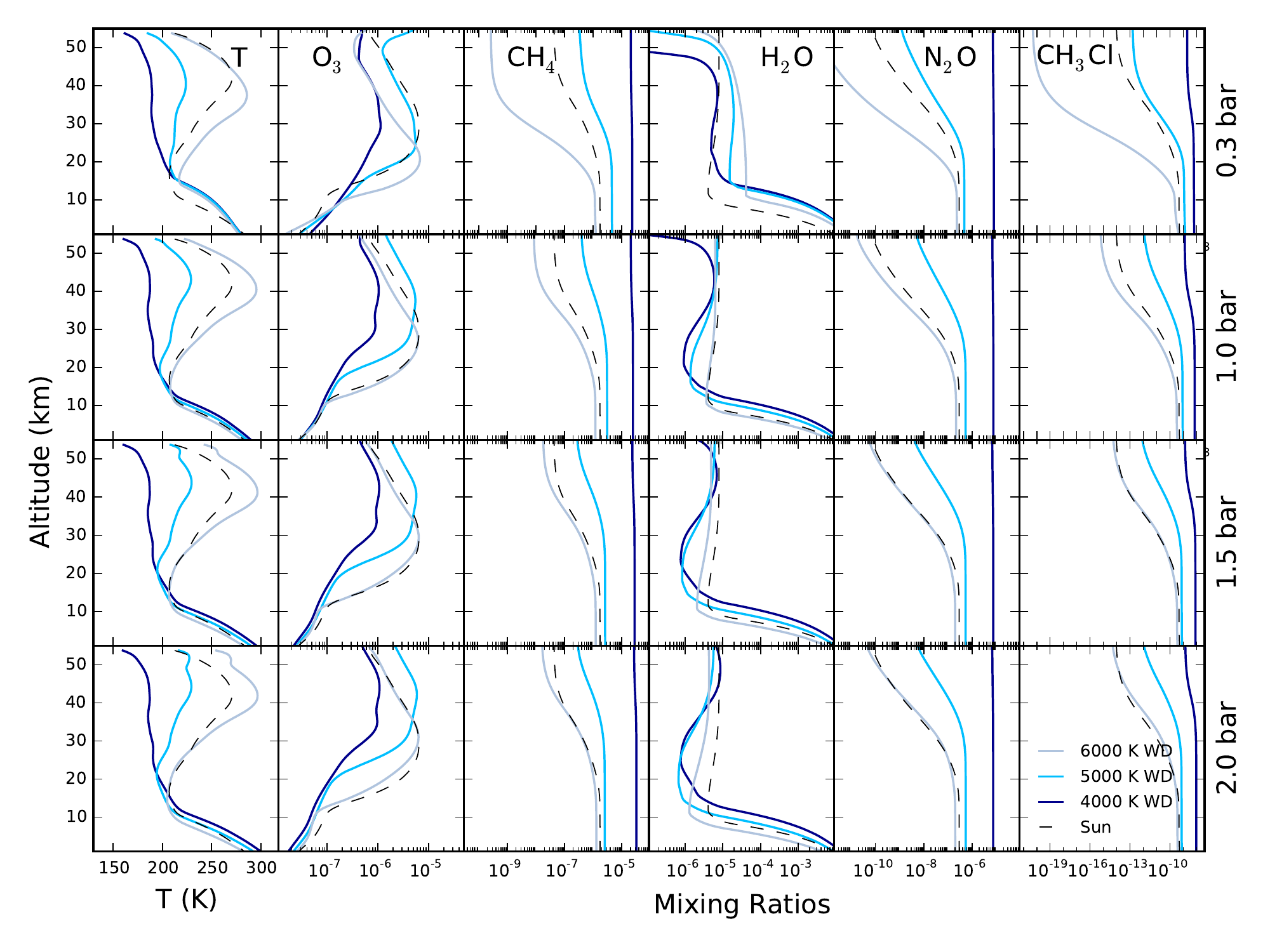}
\caption{Temperature and chemical mixing ratios for some biological interesting species for model planets orbiting a WD (blue solid lines) with 3 different surface temperatures (6000~K, 5000~K and 4000~K). Earth's values are shown for comparison (black dashed line). \label{all_photochemistry}}
\end{figure*}

\subsubsection{WD photochemistry: Earth-analogue: 1~bar surface pressure models}

Planet models with a 1~bar surface pressure, analogous to Earth, orbiting WDs show surface temperatures 285.6~K, 290.8~K, 294.3~K for WD surface temperatures of 6000~K, 5000~K and 4000~K respectively. 

However, the different WD UV environments compared to the Sun's lead to a 5\% increase in ozone column depth for the 6000~K WD surface temperature, and a 28\% and 67\% decrease of ozone column depth for the 5000~K and 4000~K WDs compared to Earth. 

Higher ozone levels cause a higher rate of the R\ref{OH} reaction, producing OH. Increased amounts of OH for higher UV levels depletes CH$_4$ via the R\ref{CH4} reaction, and similarly CH$_3$Cl via the R\ref{CH3Cl} reaction (see Figure~\ref{all_photochemistry}).

\subsubsection{WD photochemistry eroded atmospheres: 0.3~bar surface pressure models}
Planet models with a 0.3~bar surface pressure, analogous to eroded atmospheres, orbiting WDs show lower surfaces temperatures than 1.0~bar models, increasing with decreasing WD surface temperature from 280.7~K to 282.8~K, respectively. The different UV WD environments, the decreased overall amount of oxygen, as well as the decreased density of the atmosphere lead to a 33\%, 52\%, and 80\% decrease in ozone column depth compared to Earth. Longer mean-free-paths due to lower atmospheric density result in photolysis occurring at lower altitudes, causing ozone to form closer to the ground (see Figure~\ref{all_photochemistry}).

\subsubsection{WD photochemistry: Higher surface pressure planets: 1.5 \& 2.0~bar surface pressure models}
The 1.5~bar surface pressure models, for higher surface pressure planets, orbiting WDs show surfaces temperatures of 288.0~K, 295.0~K, 300.4~K for WD surface temperatures of 6000~K, 5000~K and 4000~K respectively. The different UV environments compared to the Sun and the increase in overall oxygen content as well as density of the atmosphere lead to a 20\% increase in ozone column depth for the 6000~K WD surface temperature, and a 22\% and 61\% decrease in ozone column depth compared to Earth for the 5000~K and 4000~K WDs, respectively. 

Planet models with a 2~bar surface pressure show surfaces temperatures of 289.0~K, 298.1~K, 305.1~K for WD surface temperatures of 6000~K, 5000~K and 4000~K respectively. The different UV environments compared to the Sun and the increase in overall oxygen content as well as density of the atmosphere lead to a 28\% increase in ozone column depth for the 6000~K WD surface temperature, and a17\%, and 57\% decrease in ozone column depth compared to Earth, less than for the 1.5~bar surface pressure case.

\subsection{UV surface environment around an evolving WD}
The WD cooling process provides a changing luminosity as well as UV environment at a set orbital distance, as seen in Figure~\ref{WD_models}. The UV surface environment for our planetary models from eroded atmospheres to dense atmospheres with increased surface pressure is shown in Figure~\ref{ground_UV}, with integrated fluxes for UVA (315-400~nm), UVB (280-315~nm), and UVC (121.6-280~nm) are shown in Table~\ref{UV_all}, with comparisons to the Earth-Sun system's integrated fluxes in Table~\ref{UV_all_earth}. Note that this model does not take scattering or clouds into account and thus overestimates the amount of UV that reaches the surface. However the comparison between the values and models for the UV environment on present-day Earth gives a clearer picture of the level of UV radiation that reaches the ground, compared to our own planet. High energetic UV is capable of causing harm to biological molecules, like DNA (e.g.\ \cite{voet63,diffe91,mats91,tevi93,cock98,kerw07}). Present day Earth surface life is protected by the ozone layer, which shields the surface from the most biologically dangerous radiation (UVC).

\subsubsection{Surface UV environments: Earth 1~bar surface pressure models}
For comparison we first model the amount of radiation that reaches the Earth's surface (see also \cite{rugh15}).  For our present-day Earth model the integrated  UVA ground flux compared to the UVA ITOA flux is 70\%. For integrated UVB flux, which is partially shielded by ozone, 11\% of the ITOA UVB flux reaches the surface. The UVC flux is almost completely shielded by an ozone layer, and only 5.4$\times10^{-18}$\% of the ITOA integrated UVC flux reaches the surface (see Table~\ref{ground_UV}).

\begin{sidewaystable*}[h!]
\begin{center}
\caption{UV Integrated Fluxes \label{UV_all}}
\begin{tabular}{ccccc|ccc|ccc}
Host T$_{\footnotesize \mbox{eff}}$ & Pressure & \multicolumn{3}{c}{UVA 315 - 400 nm (W/m$^2$)}  & \multicolumn{3}{c}{UVB 280 - 315 nm (W/m$^2$)} & \multicolumn{3}{c}{UVC 121.6 - 280 nm (W/m$^2$)}\\
\cline{3-5} \cline{6-11}
(K) & (bar) & ITOA & Ground & \% to ground & ITOA & Ground &\% to ground & ITOA & Ground &\% to ground  \\
\hline
Present day Earth & 1.0 & 72.4 & 50.6 & 70 & 18.9 & 2.2 & 11 & 7.1 & 3.9E-19 & 5.4E-18 \\
\hline
6000 & 	 0.3 & 	 113.2 & 	 105.1 & 	93	&	 32.4 & 	 7.4 & 	23	&	 18.5 & 	 1.9E-10 & 	 1.0E-9 \\
6000 & 	 1.0 & 	 113.2 & 	 78.9 & 	70	&	 32.4 & 	 3.5 & 	11	&	 18.5 & 	 1.9E-19 & 	 1.0E-18 \\
6000 & 	 1.5 & 	 113.2 & 	 67.8 & 	60	&	 32.4 & 	 2.5 & 	7.6	&	 18.5 & 	 3.3E-22 & 	 1.8E-21 \\
6000 & 	 2.0 & 	 113.2 & 	 59.9 & 	53	&	 32.4 & 	 2.0 & 	6.1	&	 18.5 & 	 1.8E-23 & 	 9.6E-23 \\
\hline
5000 & 	 0.3 & 	 39.4 & 	 37.6 & 	95	&	 6.6 & 	 2.0 & 	31	&	 2.3 & 	 6.6E-10 & 	 2.9E-8 \\
5000 & 	 1.0 & 	 39.4 & 	 28.6 & 	73	&	 6.6 & 	 1.0 & 	16	&	 2.3 & 	 2.5E-14 & 	 1.1E-12 \\
5000 & 	 1.5 & 	 39.4 & 	 24.6 & 	62	&	 6.6& 	 0.80 & 	12	&	 2.3 & 	 1.4E-15 & 	 6.2E-14 \\
5000 & 	 2.0 & 	 39.4 & 	 21.7 & 	55	&	 6.6 & 	 0.63 & 	9.5	&	 2.3 & 	 9.9E-17 & 	 4.3E-15 \\
\hline
4000 & 	 0.3 & 	 10.4 & 	 10.2 & 	98	&	 0.98 & 	 0.47 & 	48	&	 0.17 & 	 7.1E-6 & 	 4E-3 \\
4000 & 	 1.0 & 	 10.4 & 	 7.9 & 	76	&	 0.98 & 	 0.25 & 	26	&	 0.17 & 	 2.1E-8 & 	 1.2E-5 \\
4000 & 	 1.5 & 	 10.4 & 	 6.8 & 	65	&	 0.98& 	 0.19 & 	19	&	 0.17 & 	 1.6E-9 & 	 9.9E-7\\
4000 & 	 2.0 & 	 10.4 & 	 6.0 & 	58	&	 0.98 & 	 0.15 & 	15	&	 0.17 & 	 2.4E-10 & 1.4E-7 \\
\hline
\end{tabular}
\end{center}
\end{sidewaystable*}

\subsubsection{WD surface UV environments: Earth-analogue: 1 bar surface pressure models}
For our 1~bar surface pressure planet models, the amount of UVA flux at the surface compared to the ITOA integrated flux increases from 70\% to 76\% for the 6000~K to 4000~K WD surface temperature cases respectively, compared to the 70\% for present day Earth models. For UVB, it increases from 11\%, to 26\% of the ITOA integrated UVB for the 6000~K to 4000~K WD surface temperature cases respectively, compared to 11\% for Earth models. The UVC flux is almost completely shielded by an ozone layer, and is only 1.0$\times10^{-18}$\% to 1.2$\times10^{-5}$\% of the ITOA integrated UVC flux, for the 6000~K to 4000~K WD surface temperature cases respectively, compared to 5.4$\times10^{-18}$\% on present day Earth.

\begin{table*}[t!]
\begin{center}
\caption{UV Integrated fluxes compared to Earth \label{UV_all_earth}}
\begin{tabular}{cc|ccl}
Host T$_{\footnotesize \mbox{eff}}$ & Pressure & \multicolumn{3}{c}{WD UV/ Present day Earth UV}  \\
\cline{3-5} 
(K) & (bar) & UVA (315-400 nm) & UVB (280-315 nm) & UVC (121.6-280 nm)  \\
\hline
6000 & 	 0.3 & 	2.1	&	3.3	&	4.9$\times 10^{8}$	\\
6000 & 	 1.0 & 	1.6	&	1.6	&	4.9$\times 10^{-1}$	\\
6000 & 	 1.5 & 	1.3	&	1.1	&	8.6$\times 10^{-4}$	\\
6000 & 	 2.0 & 	1.2	&	0.90	&	4.7$\times 10^{-5}$	\\
\hline
5000 & 	 0.3 & 	0.74	&	0.90	&	1.7$\times 10^{9}$	\\
5000 & 	 1.0 & 	0.57	&	0.45	&	6.5$\times 10^{4}$	\\
5000 & 	 1.5 & 	0.49	&	0.36	&	3.6$\times 10^{3}$	\\
5000 & 	 2.0 & 	0.43	&	0.28	&	2.6$\times 10^{2}$	\\
\hline
4000 & 	 0.3 & 	0.20	&	0.21	&	1.8$\times 10^{13}$	\\
4000 & 	 1.0 & 	0.16	&	0.11	&	5.3$\times 10^{10}$	\\
4000 & 	 1.5 & 	0.13	&	0.085	&	4.2$\times 10^{9}$	\\
4000 & 	 2.0 & 	0.12	&	0.067	&	6.2$\times 10^{8}$	\\
\hline
\end{tabular}
\end{center}
\end{table*}

\subsubsection{WD surface UV environments: Eroded atmospheres: 0.3 bar surface pressure models}

For our 0.3~bar surface pressure planet models, the amount of integrated UVA surface flux compared to the ITOA integrated flux increases from 93\% to 98\% for the 6000~K to 4000~K WD surface temperature cases, {compared to 70\% for present day Earth models. For integrated UVB flux there is an increase from 23\%to 48\% of the ITOA UVB for the 6000~K to the 4000~K WD surface temperature cases, higher than the 11\% for present day Earth.   The total integrated UVC surface flux increases from 1.0$\times10^{-9}$\% to 4.3$\times10^{-3}$\% of the ITOA flux for the 6000~K to the 4000~K WD surface temperature cases, orders of magnitude higher than the 5.4$\times10^{-18}$\% for Earth.

\begin{figure*}[t!]
\begin{center}
\includegraphics[scale=0.3]{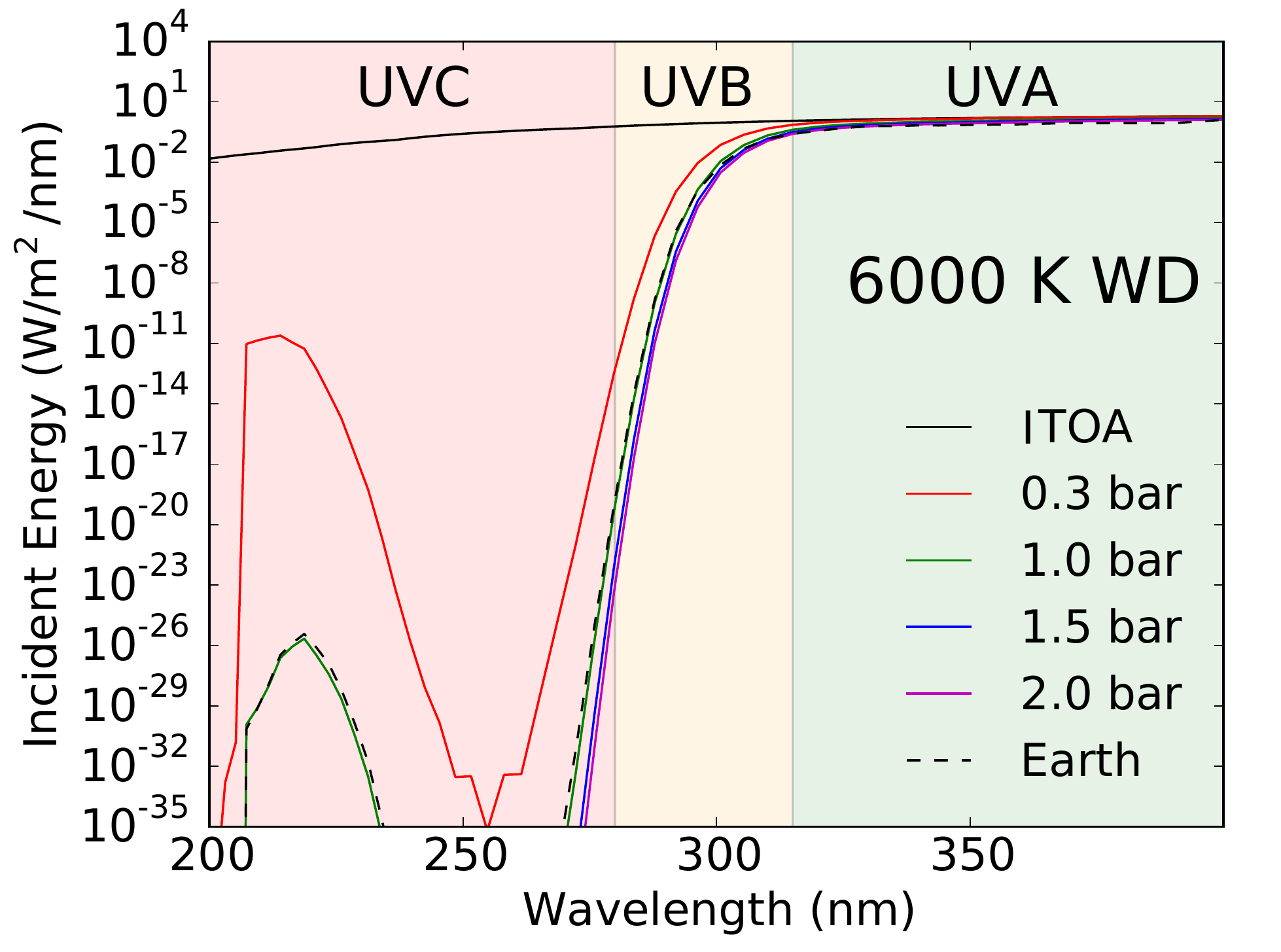}
\includegraphics[scale=0.3]{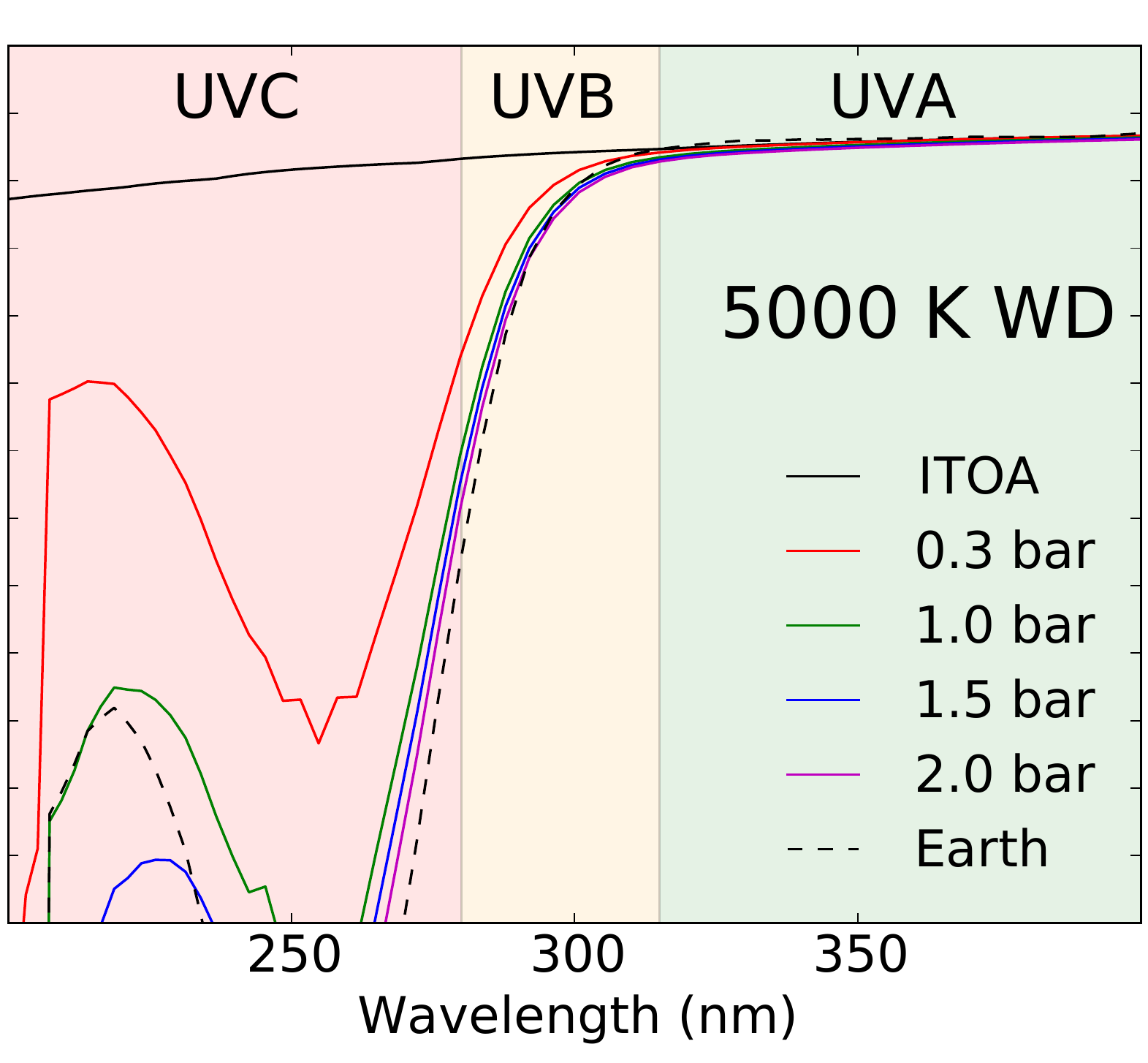}
\includegraphics[scale=0.3]{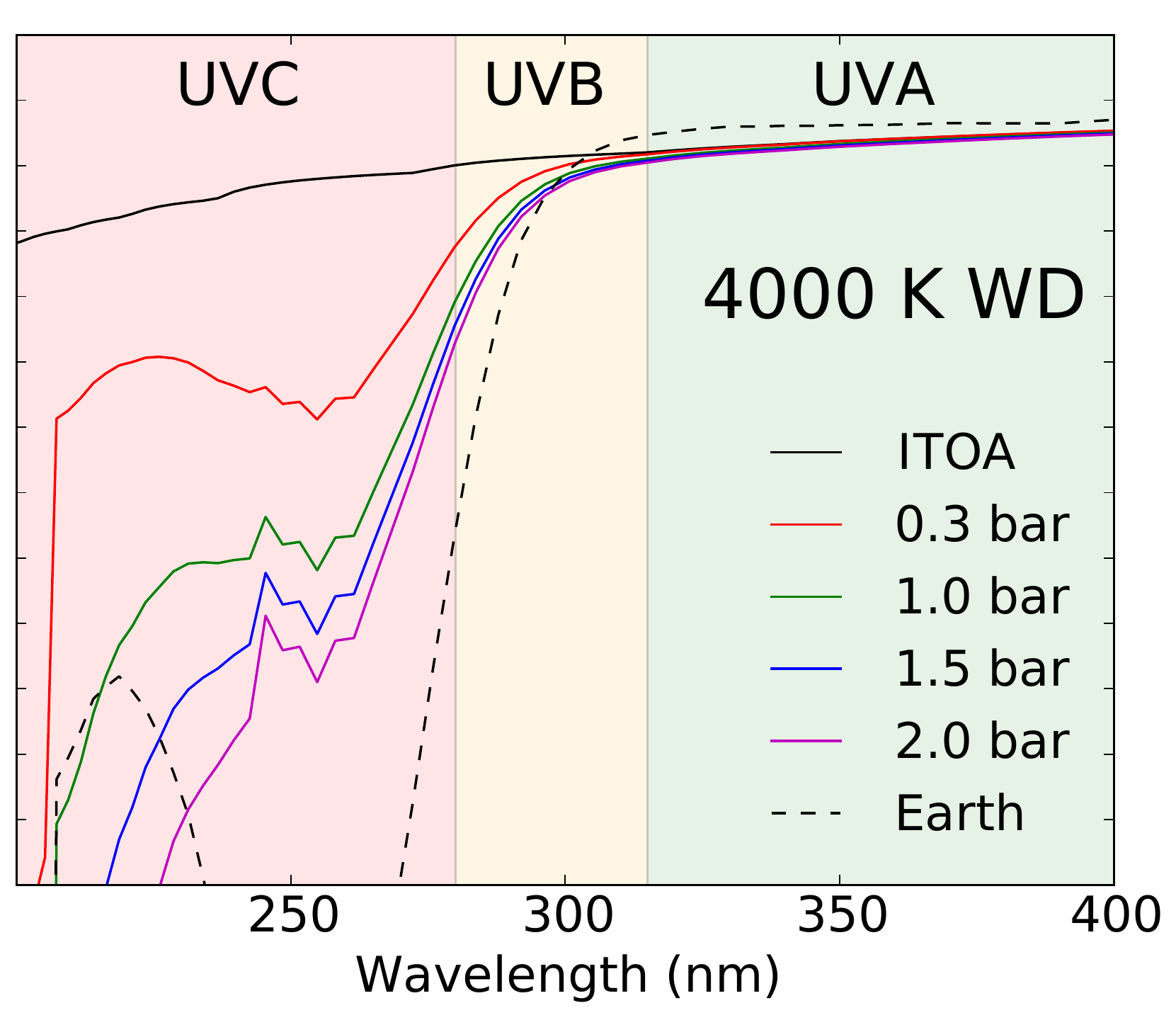}
\end{center}
\caption{Surface UV environment for planetary models with higher surface pressure (2~bar) to small planets or planets with eroded atmospheres (0.3~bar) orbiting WDs (solid lines) compared to present day Earth (dashed lines), and the incident (top-of-atmosphere) irradiation for the WD (black solid line). \label{ground_UV}}
\end{figure*}

\subsubsection{WD surface UV environments: higher surface pressure planets: 1.5~bar and 2~bar surface pressure models}

For our 1.5~bar surface pressure planet models, the amount of integrated surface UVA flux compared to the ITOA integrated flux increases from 60\% to 65\% for the 6000~K through 4000~K WD surface temperature cases, less than the 70\% for present day Earth models. For UVB surface flux there was an increase of 7.6\% to 19\% of the ITOA integrated UVB flux for the 6000~K to 4000~K WD surface temperature cases, comparable to 11\% for Earth models. The UVC ground flux increased from 1.8$\times10^{-21}$\% to 9.9$\times10^{-7}$\% of the ITOA integrated flux for the 6000~K to the 4000~K WD surface temperature cases, respectively, compared to 5.4$\times10^{-18}$\% for present day Earth.

For our 2~bar surface pressure planet models, the amount of integrated UVA ground flux compared to the ITOA integrated flux increases from 53\% to 58\% for the 6000 K to 4000 K WD surface temperature cases, respectively, less than 70\% for present day Earth models. For UVB integrated ground flux there is an increase of 6.1\% to 15\% of ITOA UVB integrated flux reaching the surface for the 6000 K through 4000 K WD surface temperature cases respectively, comparable to 11\% for present day Earth.  The integrated UVC flux to the ground increases from 9.6$\times10^{-23}$\% to 1.4$\times10^{-7}$\% of the original integrated ITOA flux for the 6000~K to 4000~K WD surface temperature cases, respectively, compared to 5.4$\times10^{-18}$\% for Earth models.

The planetary models with the 4000~K WD surface temperature host show the highest overall UVC ground flux despite having a lower UVC ITOA integrated flux, because of the lower ozone level in these atmospheres, compared to hotter WD models.

Overall, Figure~\ref{ground_UV} shows that the UV surface environment for our model planets orbiting WDs. The 1~bar surface pressure model for a 6000~K WD surface temperature model receives similar UV surface levels as present day Earth models. Only the 1.5~bar and 2~bar surface pressure models for the same 6000~K WD surface temperature receive a lower UV integrated surface flux than present day Earth. For all other models, the UV surface flux is higher than for present day Earth, especially the UVC environment (see Table~\ref{UV_all_earth}).

\subsection{Planetary environments for planets in the HZ through the evolution of a WD}

We model two case studies, A and B, as shown in Figure~\ref{WD_HZ}, which explore the environment of a planet with a 1~bar surface pressure (i.e., an Earth analogue) in the HZ of a WD during its evolution. Case~A shows the maximum time a planet can stay in the HZ of the WD as the WD cools from 6000~K to 4000~K. This amounts to $\sim$6~Gyr for the classical (conservative) HZ and $\sim$8.5~Gyr for the empirical HZ. Case~B focuses on a planet that initially receives the same irradiance as Earth around a WD with a surface temperature of 5000~K.  As the WD cools from 6000~K to 4000~K the planet spends $\sim$4~Gyr in the conservative HZ and $\sim$7~Gyr in the empirical HZ. Model parameters and results are shown in Tables~\ref{UV_cases} and~\ref{UV_evol}, with comparisons to Earth in Table~\ref{UV_evol_earth} for both cases. UV environments are shown in Figure~\ref{HZ_caseA} for Case~A and Figure~\ref{HZ_caseB} for Case~B, and photochemistry is shown in Figure~\ref{evol_photochemistry}. 

Such a planet could have orbited in the HZ of a cool WD for longer than the Earth has existed. Single-celled life likely emerged on Earth less than 1~Gyr after its formation, with multicellular life following 2.7~Gyr later.

\begin{table*}[t!] 
\begin{center}
\caption{Case A \& Case B Results\label{UV_cases}}
\begin{tabular}{c|ccc|ccc}
\hline
  &    \multicolumn{3}{c}{Case A: r = 0.0085 AU} & \multicolumn{3}{c}{Case B: r = 0.0069 AU} \\
\cline{2-7}
Host & S$_{eff}$ & Surface & Ozone Column & S$_{eff}$ & Surface & Ozone Column \\
(K) & & T$_{\footnotesize \mbox{eff}}$ (K) & Depth (cm$^{-2}$) &  & T$_{\footnotesize \mbox{eff}}$ (K) & Depth (cm$^{-2}$) \\
\hline
6000	&	1.3	&	328.4	&	1.7$\times 10^{18}$ 	&	2.1	&	RG$^*$	&	RG$^*$\\
5000	&	0.64	&	249.6	&	4.4$\times 10^{18}$ 	&	1.0	&	290.8	&	3.3$\times 10^{18}$   \\
4000	&	0.26 &	191.9	&	9.6$\times 10^{17}$ 	&	0.41	&	216.9	&	1.3$\times 10^{18}$  \\
\hline
\end{tabular}
\end{center}
$^*$RG indicates a runaway greenhouse state
\end{table*}

\subsubsection{Case A}
The orbital distance of 0.0085 AU or 1.3~million km leads to a changing illumination by the cooling WD from S$_{eff}$ of 1.34 for a 6000~K WD surface temperature, to S$_{eff}$ of 0.64 for a 5000~K WD surface temperature, to S$_{eff}$ of 0.26 for the 4000~K WD surface temperature model. This leads to decreasing planetary surfaces temperatures of 328.4~K, 249.6~K, 191.9~K for WD surface temperatures of 6000~K, 5000~K and 4000 K, respectively. 

In our model the planet's surface temperature is on average above freezing for only a part of the time the planet spends in the WD's HZ, a few billion years (Figure~\ref{WD_cooling} and Figure~\ref{WD_HZ}). However we have not adjusted the CO$_2$ content in our models, thus if a cycle similar to the carbonate silicate cycle on Earth existed on such a planet, CO$_2$ concentration should increase, heating the planet's surface temperature and keeping it from freezing. As shortly discussed, a geologically active planet is the underlying assumption of continued habitability during a star's evolution, and is what led to the concept of the HZ (see e.g.\ \cite{kast93}). For a WD, a similar cycle could be possible, extending the time the surface of such a planet can be above freezing to the full range of its time in the WD's HZ.

The different WD UV environments compared to the Sun lead to a 29\%, 18\%, and 82\% decrease in overall ozone column depth compared to Earth. As shown in Figure~\ref{evol_photochemistry}, ozone levels are highest for a WD model of 5000~K, providing a slightly higher shielding from UV than present day Earth. However, during both the 6000~K and 4000~K WD surface temperature stages, the UVC flux at the surface increases substantially for different reasons. During the 6000~K WD surface temperature period, UV photons with $\lambda<$~320 [R\ref{O3}]  are available. These can dissociate ozone, therefore increasing the UV surface flux. The 4000~K WD model has a comparably high amount of surface UVC radiation because low UV irradiation from the 4000~K WD cannot initially produce enough ozone to efficiently shield the surface of the planet. 

The amount of UVA integrated surface flux compared to the ITOA integrated flux is 71\%, 74\%, and 78\%, for the 6000~K, 5000~K, and 4000~K WD surface temperature cases, comparable to the 70\% for model Earth models. For UVB flux, which is partially shielded by ozone, 24\%, 15\%, and 32\% of the integrated ITOA UVB reach the surface, for the 6000~K, 5000~K, and 4000~K WD surface temperature cases, compared to 11\% for present day Earth models. The  UVC flux is almost completely shielded by an ozone layer, and the percentage of the integrated ITOA flux reaching the planetary surface  1.0$\times10^{-5}$\%, 2.4$\times10^{-14}$\%, and 1.0$\times10^{-3}$\%, for the 6000~K, 5000~K, and 4000~K WD surface temperature cases, respectively, compared to 5.4$\times10^{-18}$\% for present day Earth models.

\begin{figure*}
\begin{center}
\includegraphics[scale=0.4]{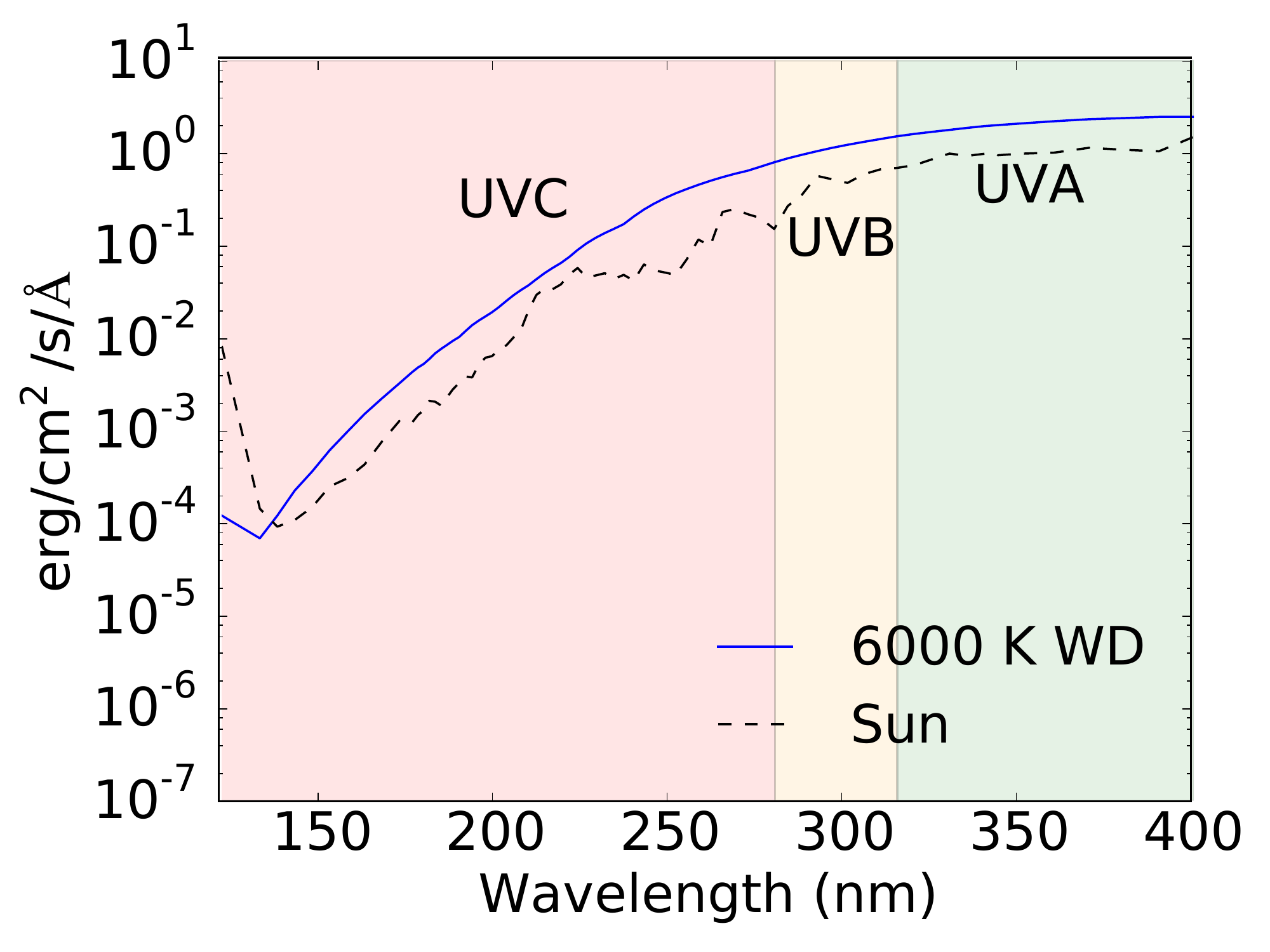}
\includegraphics[scale=0.40]{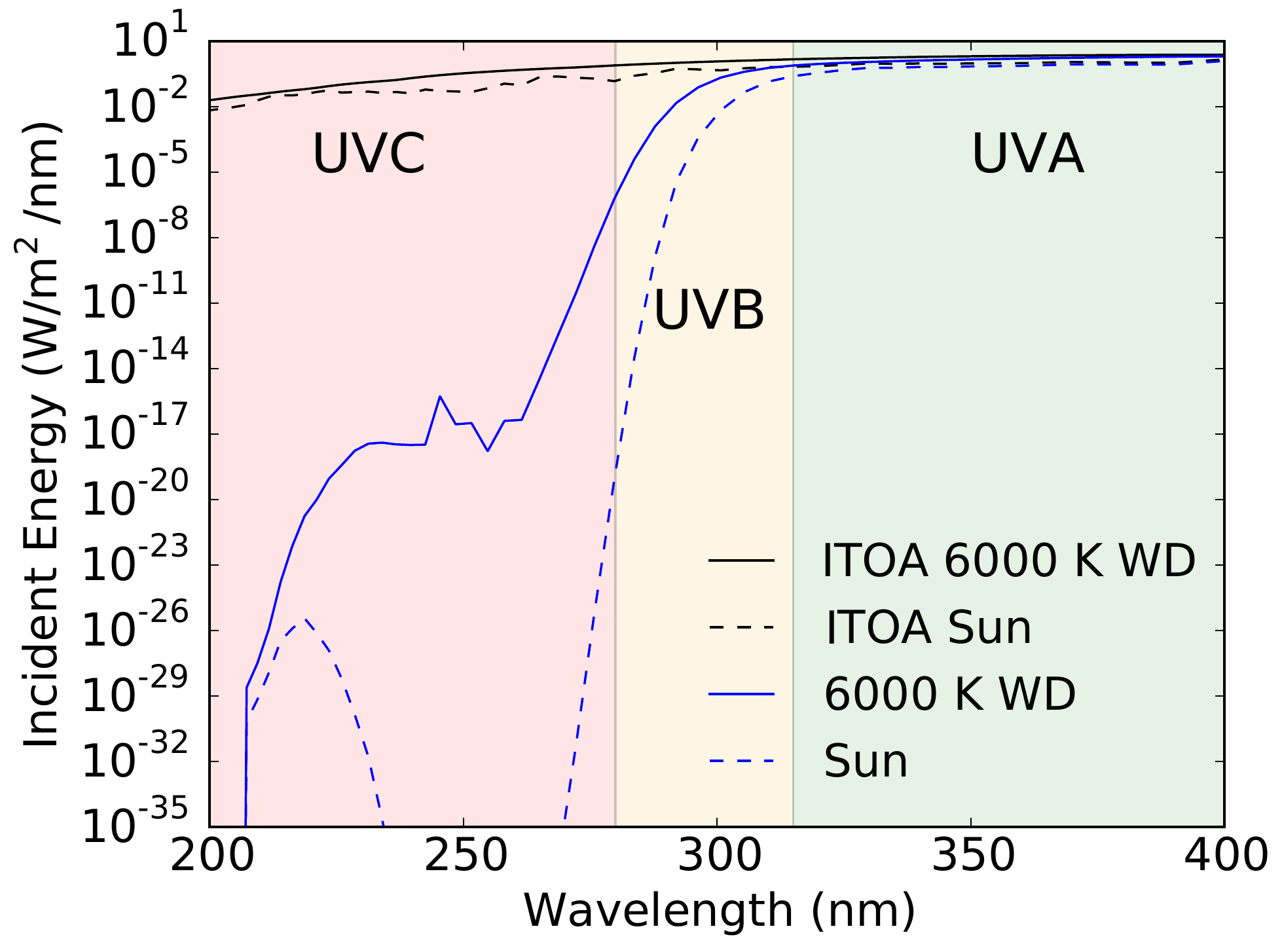} \\
\includegraphics[scale=0.4]{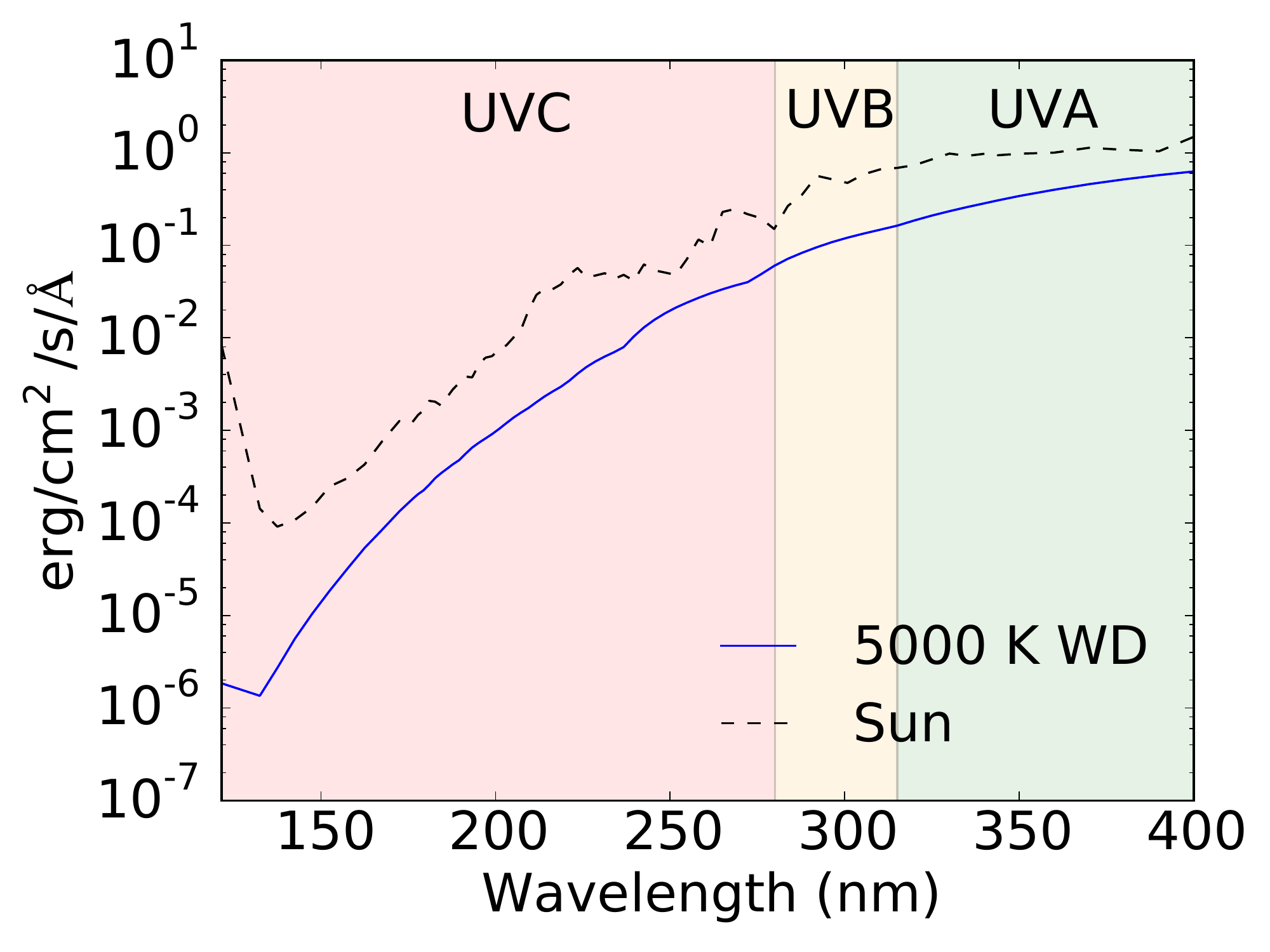}
\includegraphics[scale=0.4]{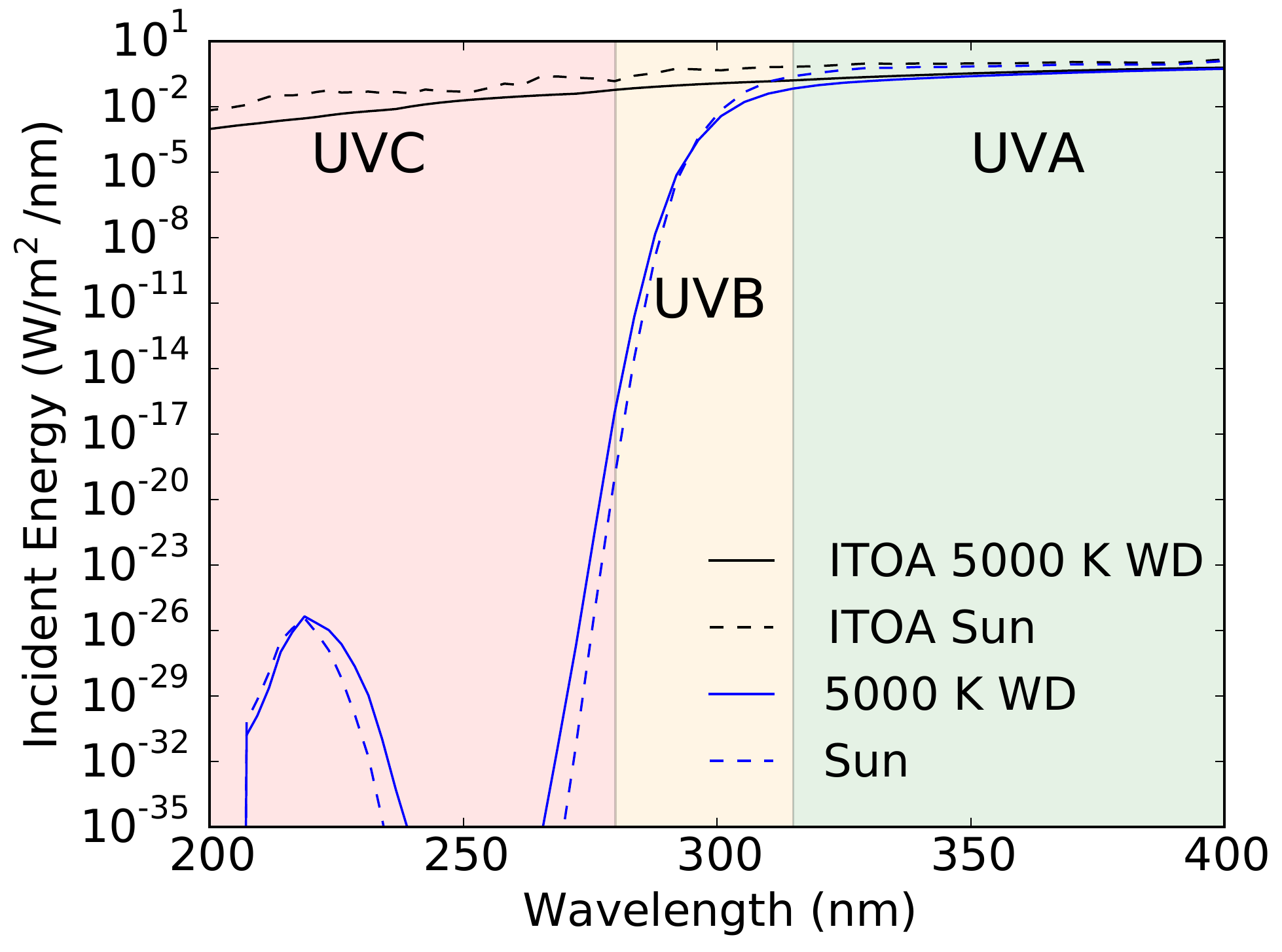}\\
\includegraphics[scale=0.4]{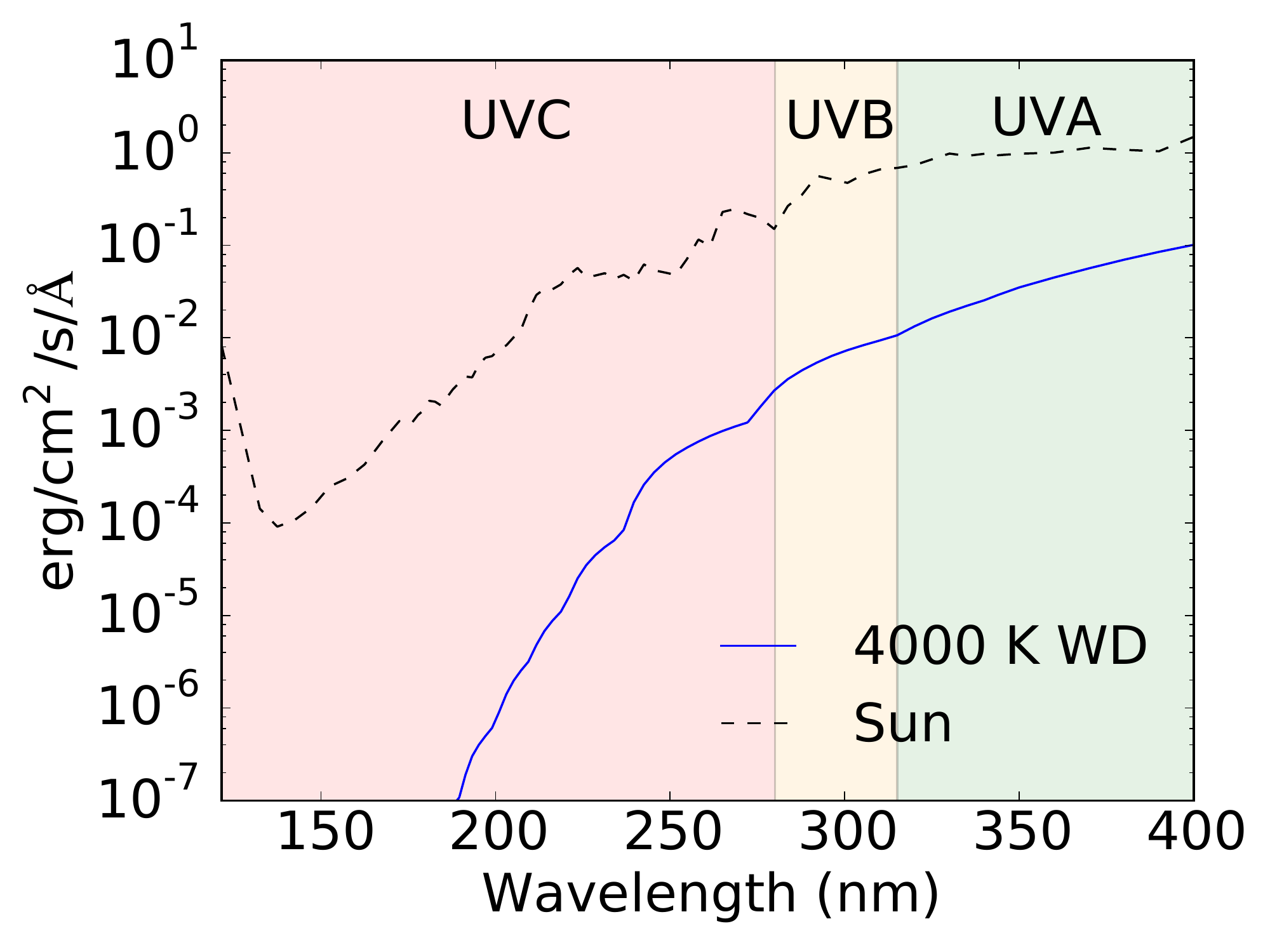}
\includegraphics[scale=0.4]{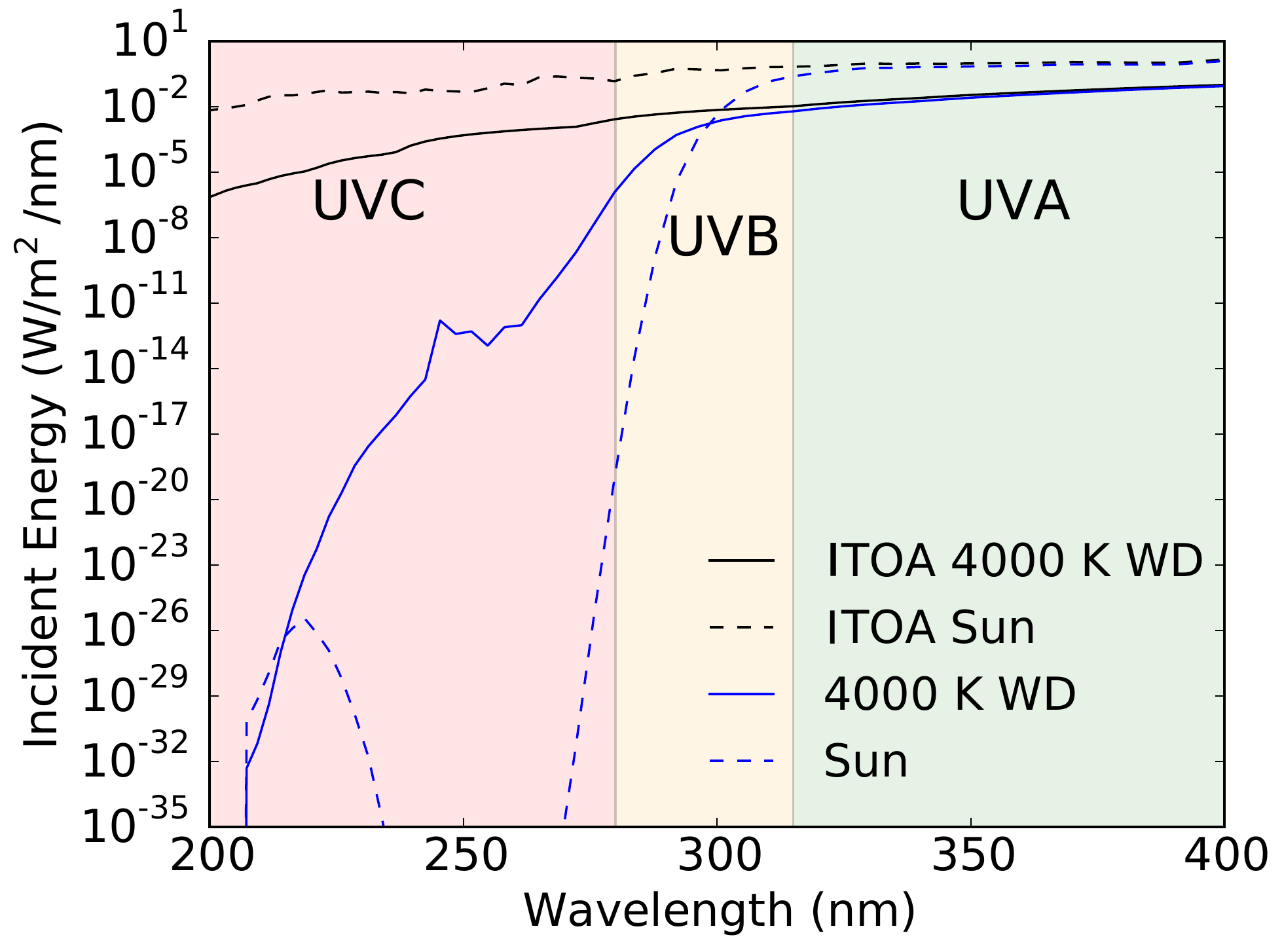}
\end{center}
\caption{UV surface flux for Case~A with a planet orbiting at 0.0085~AU from its host, corresponding to S$_{eff}$~=~1.34 for for the 6000~K model, S$_{eff}$~=~0.64 for a 5000~K model, and S$_{eff}$~=~0.26 for 4000~K models.  \label{HZ_caseA}}
\end{figure*}

\begin{sidewaystable*}[h!]
\begin{center}
\caption{UV Integrated fluxes evolution \label{UV_evol}}
\begin{tabular}{ccccc|ccc|ccc}
Case & Host T$_{\footnotesize \mbox{eff}}$ & \multicolumn{3}{c}{UVA 315 - 400 nm (W/m$^2$)}  & \multicolumn{3}{c}{UVB 280 - 315 nm (W/m$^2$)} & \multicolumn{3}{c}{UVC 121.6 - 280 nm (W/m$^2$)}\\
\cline{3-5} \cline{6-11}
 & (K) & ITOA & Ground & \% to ground & ITOA & Ground &\% to ground & ITOA & Ground &\% to ground  \\
 \hline
Present day Earth & 5750 & 72.4 & 50.6 & 70 & 18.9 & 2.2 & 11 & 7.1 & 3.9$\times10^{-19}$ & 5.4$\times10^{-18}$ \\
\hline
A & 6000 &151.3& 107.0 & 71 & 43.4 & 10.2 & 24 & 24.7 & 2.51$\times10^{-6}$ & 1.0$\times10^{-5}$ \\
A & 5000 & 25.4& 18.9 & 74 & 4.3 & 0.64 & 15 & 1.5 & 3.6$\times10^{-16}$ & 2.4$\times10^{-14}$ \\
A & 4000 & 4.3 & 3.3 & 78 & 0.40 & 0.13 & 32 & 6.8$\times10^{-2}$ & 7.0$\times10^{-7}$ & 1.0$\times10^{-3}$ \\
\hline
B & 5000& 	 39.4 & 	 28.6 & 	73	&	 6.6 & 	 1.0 & 	16	&	 2.3 & 	 2.5$\times10^{-14}$ & 	 1.1$\times10^{-12}$ \\
B & 4000 & 2.8 & 2.1 & 77 & 0.26 & 9.2$\times10^{-2}$ & 36 & 4.4$\times10^{-2}$ & 5.0$\times10^{-6}$ & 1.1$\times10^{-2}$ \\
\hline
\end{tabular}
\end{center}
\end{sidewaystable*}

\subsubsection{Case B}
The orbital distance of the planet is 0.0069~AU in Case~B. This corresponds to an illumination of 1.0~S$_{eff}$ for the 5000~K WD, which evolves to 0.41~S$_{eff}$ as the WD cools to 4000~K.  For Case B planet models the surfaces temperatures are 290.8~K and 216.9~K for WD surface temperatures of 5000~K and 4000~K, respectively. The different UV environments compared to the Sun lead to a 39\% and a 77\% decrease in overall ozone column depth for WD surface temperatures of 5000~K and 4000~K, respectively, compared to Earth.

The amount of UVA ground flux compared to the ITOA integrated flux is 73\% and 77\%, for the 5000~K and 4000~K WD surface temperature models, respectively (see Figure~\ref{HZ_caseB}), comparable for the 70\% for present day Earth models. For UVB flux, which is partially shielded ozone, only 16\%, and 36\% of the integrated ITOA UVB reaches the surface, for the 5000~K and 4000~K WD surface temperature cases, compared to 11\% for present day Earth models. The UVC flux is almost completely shielded by an ozone layer, and is 1.1$\times10^{-12}$\%, and then 1.1$\times10^{-2}$\% of the ITOA integrated UVC flux, compared to 5.4$\times10^{-18}$\% for present day Earth models.

As the UV levels from the WD decrease, less ozone is produced in the planet's atmosphere and thus UVB and UVC surface levels increase. Average surface temperature decreases below freezing for outgassing rates similar to present day Earth. However, as discussed for Case~A, if a carbonate-silicate cycle exists, increased amounts greenhouse gases should increase surface temperatures above freezing for an Earth-like planet even with decreasing illumination by the cooling WD.

\begin{table*}
\begin{center}
\caption{UV Integrated fluxes evolution Earth comparison \label{UV_evol_earth}}
\begin{tabular}{cc|ccl}
Case &Host T$_{\footnotesize \mbox{eff}}$  & \multicolumn{3}{c}{WD UV/Present day Earth UV}  \\
\cline{3-5} 
 &(K) & UVA & UVB & UVC  \\

 \hline
A	&	6000	&	2.1	&	4.6	&	6.5$\times10^{12}$	\\
A	&	5000	&	3.7$\times10^{-1}$	&	2.9$\times10^{-1}$ &	9.3$\times10^{2}$	\\
A	&	4000	&	6.6$\times10^{-2}$	&	5.8$\times10^{-2}$	&	1.8$\times10^{12}$	\\
\hline
B	&	5000	&	5.7$\times10^{-1}$	&	4.5$\times10^{-1}$		&	6.48$\times10^{4}$	\\
B	&	4000	&	4.2$\times10^{-2}$	&	4.1$\times10^{-2}$		&	1.3$\times10^{13}$	\\
\hline
\end{tabular}
\end{center}
\end{table*}

\begin{figure*}[b!]
\begin{center}
\includegraphics[scale=0.4]{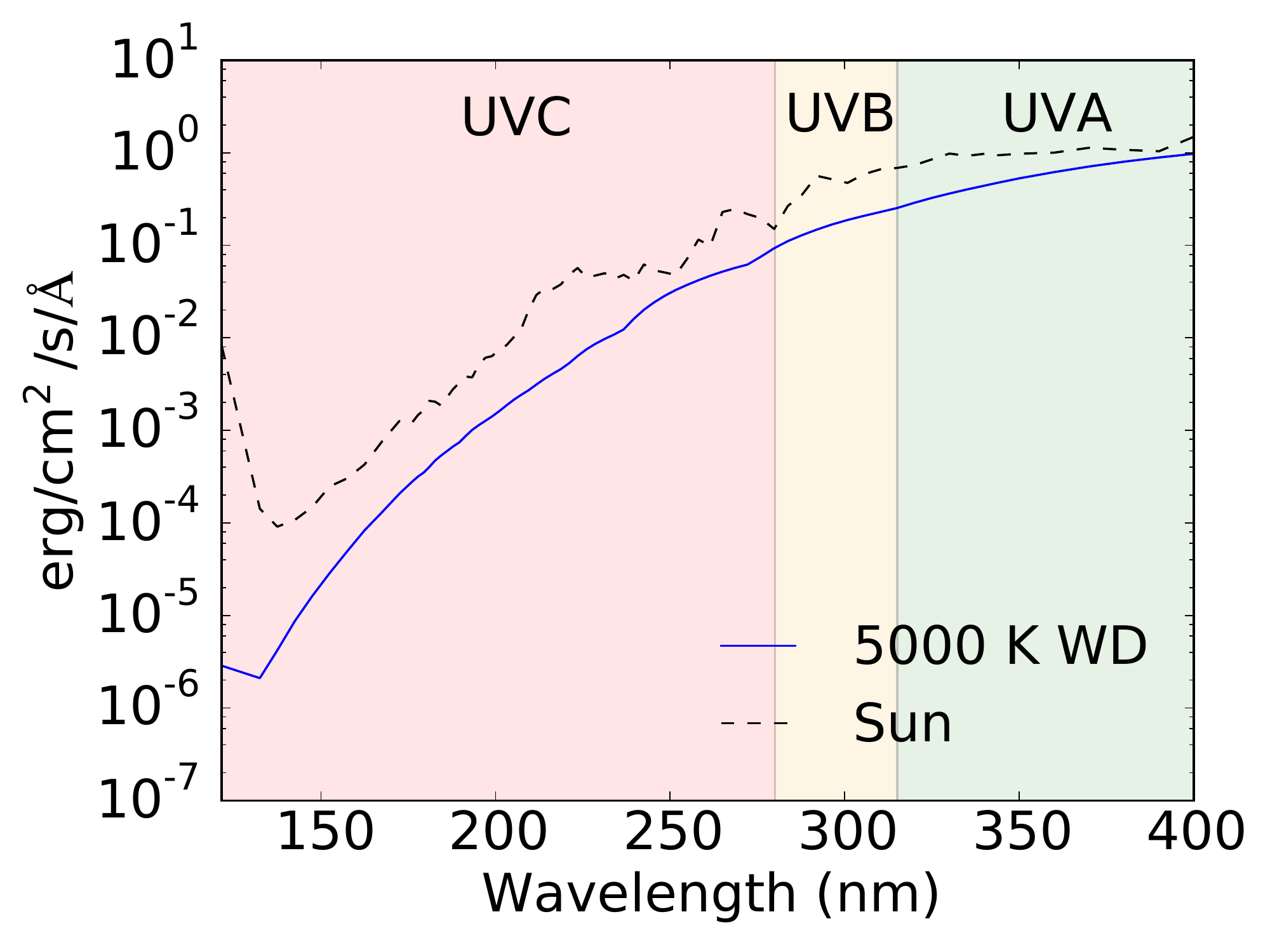}
\includegraphics[scale=0.4]{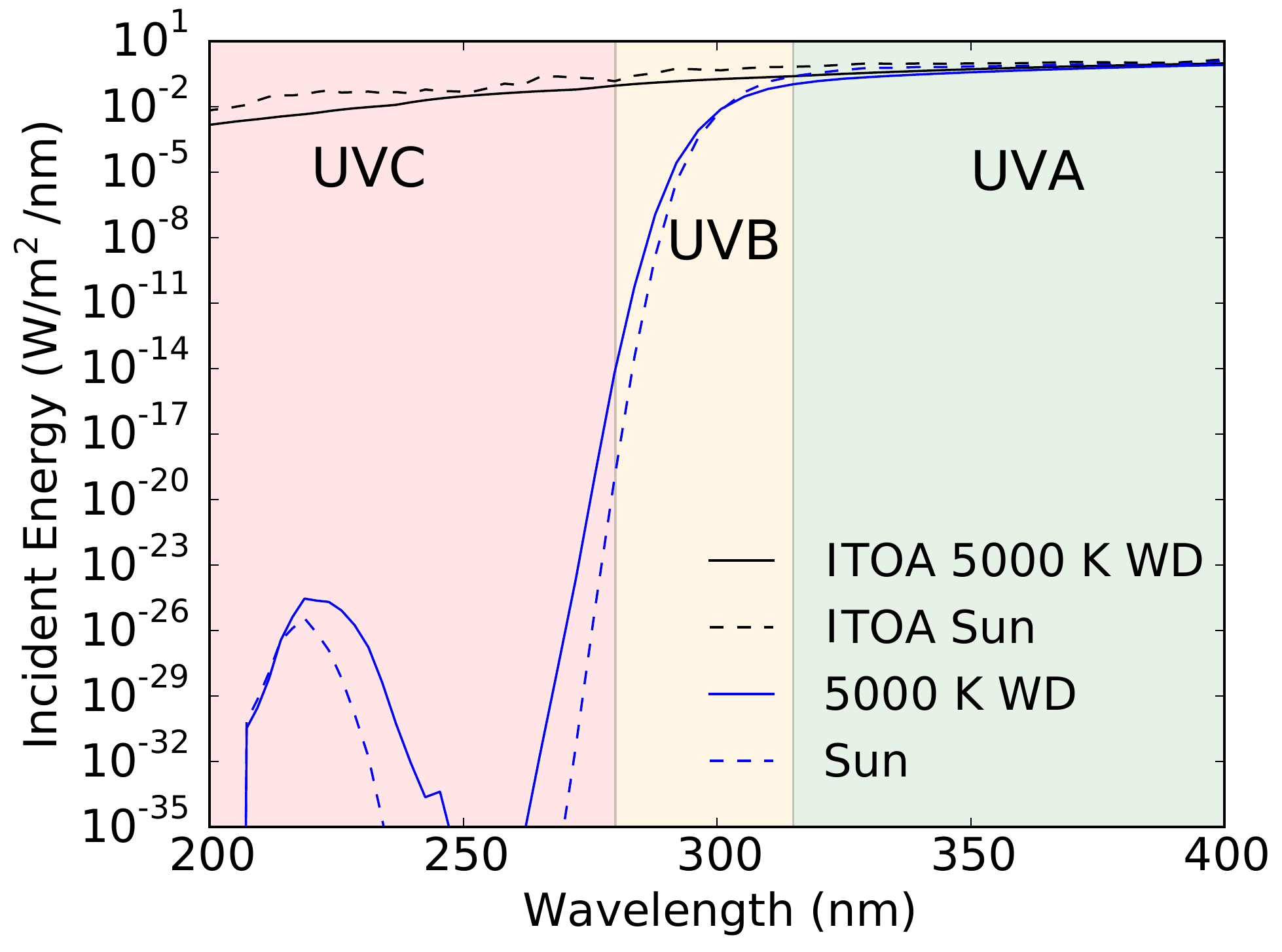}\\
\includegraphics[scale=0.4]{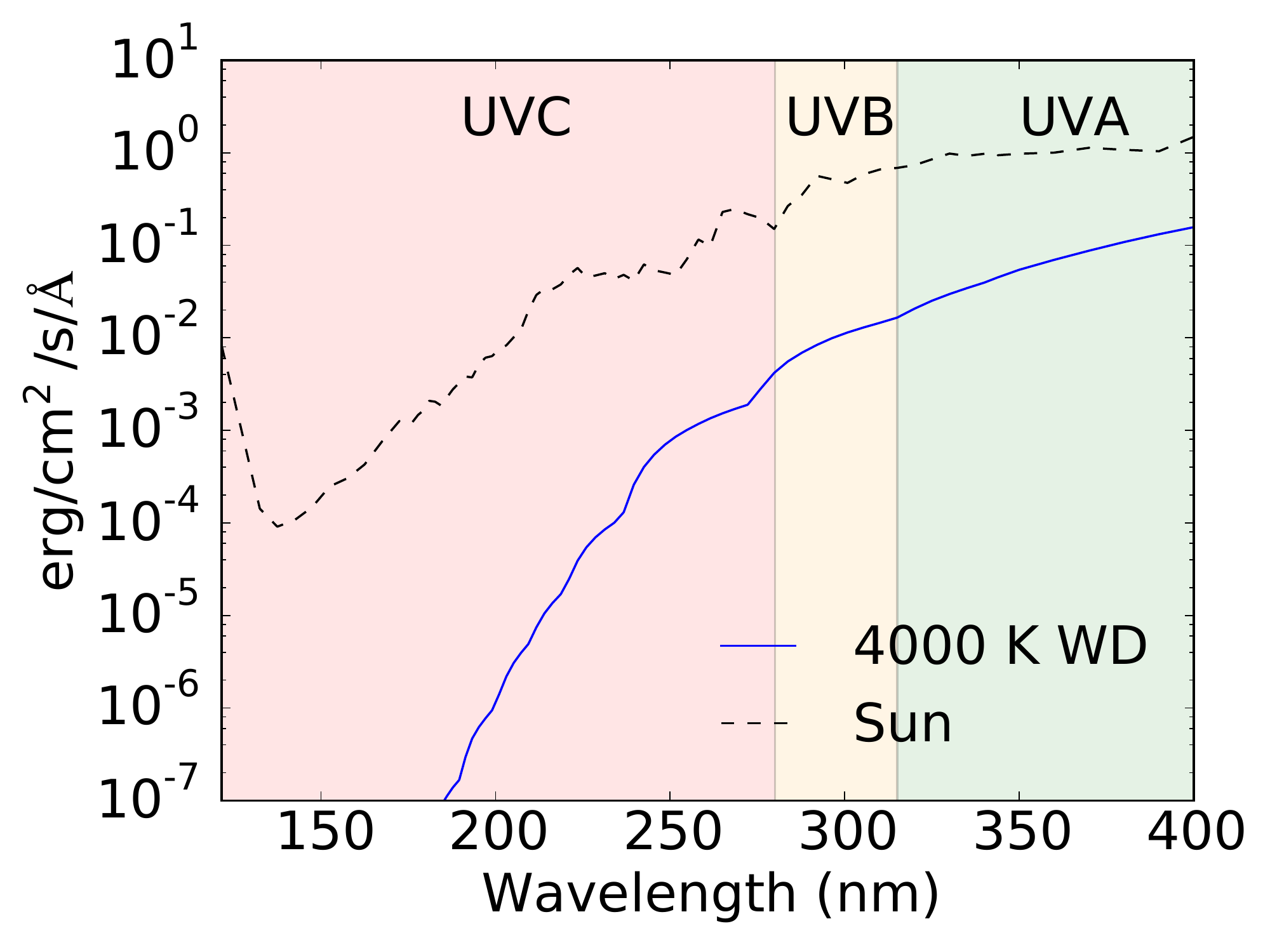}
\includegraphics[scale=0.4]{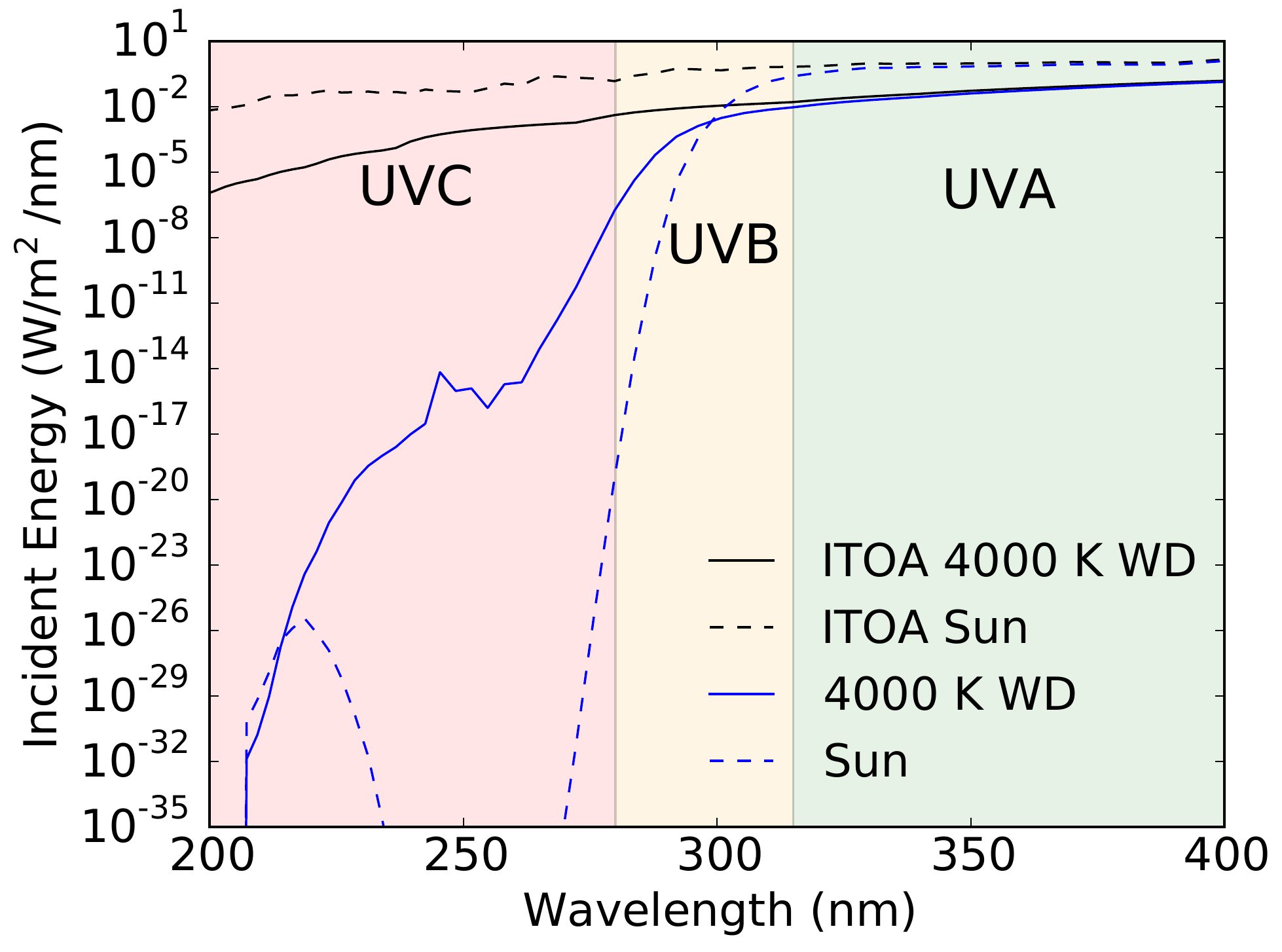}
\end{center}
\caption{UV surface flux for Case B: a planet orbiting at 0.0069~AU, corresponding to S$_{eff}$ = 1.0 for the 5000~K WD surface temperature, and S$_{eff}$ = 0.41 for the 4000~K WD surface temperature model. At 6000~K the S$_{eff}$ equals 2.07, placing the planet outside of both the conservative as well as empirical HZ.\label{HZ_caseB} }
\end{figure*}

\begin{figure*}[t!]
\begin{center}
\includegraphics[scale=0.8]{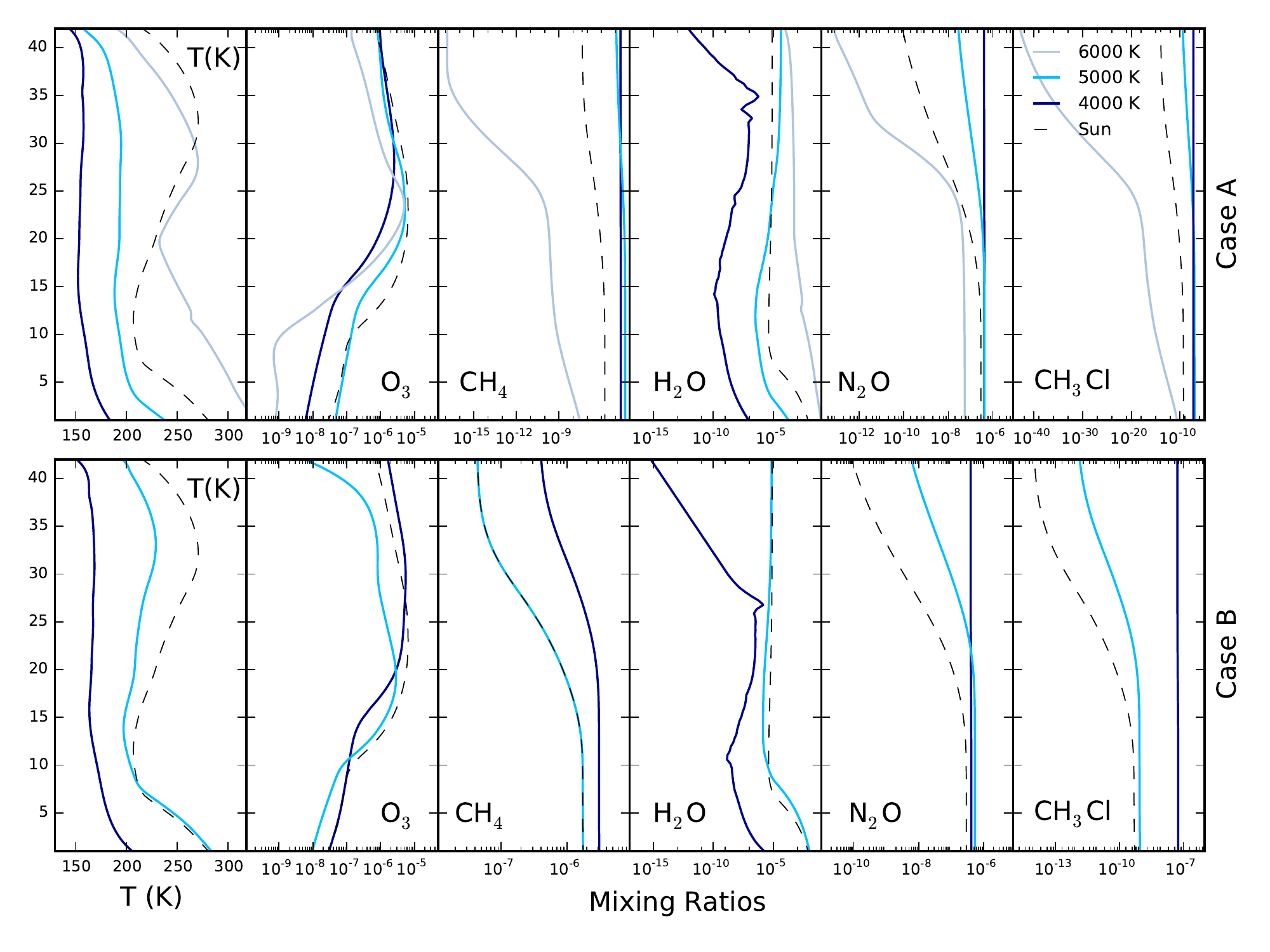}
\caption{Temperature and photochemistry profiles for Case A and B (solid lines), with the Earth-Sun profile for comparison (dashed). \label{evol_photochemistry}}
\end{center}
\end{figure*}

\section{Discussion \label{discussion}}

\subsection{How could white dwarf planets form?}
The mechanisms required for a WD planetary system to form as secondary generation objects from a disk or to survive post-main sequence evolution are not well understood. During the post main sequence evolution of the host into a WD, inner rocky planets (within 1-2~AU) would likely be destroyed (e.g.\ \cite{villa11,kuni11,villa12,must12,villa14}), while stellar mass loss would cause semimajor axis expansion for outer planets (e.g.\cite{vera16,rami16}). Exomoons of outer planets could potentially migrate inward \citep{vera16,rami16}, although the estimated occurrence rate wouldn't explain the rate of polluted WDs \citep{slui17,vane17}. Second-generation planets could even form from fall-back of debris initially expelled during the post-main sequence evolution of the host star \citep{vera16}. An interesting system is WD1145+017, a WD orbited by a rocky minor planet undergoing tidal disintegration at the system's Roche limit \citep{vand15}.

\subsection{Only dry planets? White dwarfs start with an extremely hot phase}

With no stellar mass loss during the WD phase (and assuming rapid tidal circularization of its orbit), a planet should remain at the same orbital distance during a WD's evolution. Due to the extreme change in luminosity early in a WD's evolution (Figure~\ref{WD_cooling}) a temperate planet around a cool WD would have thus experienced an extremely high luminosity and corresponding surface temperature early in its history. This would initiate a runaway greenhouse process and extreme water loss on such planets early in their history. Similarly for planets orbiting pre-main sequence M-type stars, which also are more luminous initially and should be able to initiate a runaway greenhouse phase and water loss on a temperate planet that can be found in the main sequence HZ later on (see \cite{rami14,barn15}), there may be methods of late water delivery occurring in a WD system after to the WD has cooled to a surface temperature that maintains a slower changing luminosity (e.g.\ \cite{jura10,fari13,mala16,mala17a,mala17b}).

Any Earth-like planet that is orbiting too close to the WD would remain in a runaway greenhouse stage for a certain time, until it entered the WD HZ. Whether or not such planets would have lost all their water at that point would depend strongly on the initial water reservoir, and whether it can be replenished. Thus, whether a WD planet can host water and an Earth-like atmosphere, is an open question and will depend on its evolution as well as when the WD planet is formed, from what materials it is made of and whether the possibility of a continuous water delivery exists in WD planetary systems. 

\subsection{A white dwarf planet's evolution differs from that of Earth}

A planet orbiting a WD will receive decreasing overall energy from its host during the WD's cooling process. Its evolution is very different from that of a planet orbiting a main sequence star, whose luminosity increases with time, pushing the HZ to wider separations. For a planet around a main sequence star, the increase in greenhouse gases at the inner edge of the HZ coupled with the increasing luminosity of the host star limit the time it can remain habitable. If a cycle similar to the carbonate silicate cycle could also operate on WD planets, planets on the inner edge of a WD could build up substantial greenhouse gas amounts in their atmospheres as well; however due to the decreasing luminosity of their WD host, such build up of greenhouse gases could help to maintain warm surface temperature as the WD flux decreases, possibly extending the length of time life could survive on WD planets.

\section{Conclusions  \label{conclusions}}

Our models explore the atmospheric environments of planets orbiting WDs, taking into consideration the changing surface temperatures and UV environments of WDs during their cooling process. 

We model the atmospheric composition as well as the UV surface environments of Earth-like planets orbiting WDs during different points throughout a WD's evolution. Our planet models have surface pressures ranging from 2~bar to 0.3~bar, including Earth-analog planets with 1~bar atmospheres, as well as planets with 0.3~bar surface pressure (e.g.\ eroded atmospheres) and planets with higher surface pressures of 1.5~bar and 2~bar (e.g,\ super-Earths). 

The integrated overall ozone column depth is less than on present day Earth for all our model runs, except for the models with surfaces pressures of 1~bar or above for the WD 6000~K surface temperature model, which provides similar UV to the Sun. The UV surface environment on a planet is controlled primarily by the incoming irradiation and by the planetary atmospheric composition. The UV surface environment for all our planetary models orbiting WDs show increased surface UVC flux up to several orders of magnitude compared to present day Earth, except for the models with surface pressures of 1~bar or above for the 6000~K WD model (see Table~\ref{ground_UV}). The UVC can become substantially higher for the cool 4000~K WD surface temperature model runs, making those model planet surface environments harsh for life as we know it. 

In addition to individual models that represent a planet orbiting a WD at a certain time during the cooling process (i.e., specific WD surface temperatures), we also model two planets through their evolution while the WD cools, representing two possible tracks through a WD's HZ. Both cases show a decrease of ozone for the three points modeled (for WD surface temperature of 6000~K, 5000~K and 4000~K and the corresponding irradiation on the planets) as well as an increase in UVC surface flux over time. 

Due to the extreme change in luminosity early in a WD's evolution a temperate planet around a cool WD would have experienced extremely high luminosity and corresponding surface temperature early in its history, what should initiate a runaway greenhouse process and extreme water loss on such planets early in their history. Whether or not such planets would have lost all their water at that point would depend strongly on the initial water reservoir, and whether it can be replenished, leaving open the question about whether an Earth-like planet could survive around WDs if formed. However due to the extremely favorable size ratio of an Earth-like planet compared to a WD, as well as the low luminosity of a WD compared to a main sequence host star, WD exoplanets will make interesting targets for characterization.

\acknowledgments

LK and TK acknowledge support by the Simons Foundation (SCOL \# 290357, Kaltenegger) and the Carl Sagan Institute.  We thank Piotr Kowalski for kindly providing the model spectra of cool WDs used here.


\begin{thebibliography}{}

\bibitem[Agol(2011)]{agol11} Agol, E.\ 2011, \apjl, 731, L31

\bibitem[Barnes et al.(2015)]{barn15} Barnes, R., Meadows, V.~S., \& Evans, N.\ 2015, \apj, 814, 91

\bibitem[Benvenuto \& Althaus(1999)]{ba1999} Benvenuto, O.~G., \& Althaus, L.~G.\ 1999, \mnras, 303, 30 

\bibitem[Bergeron et al.(2001)]{berg01} Bergeron, P., Leggett, S.~K., \& Ruiz, M.~T.\ 2001, VizieR Online Data Catalog, 213

\bibitem[Bergeron et al.(1997)]{berg97} Bergeron, P., Ruiz, M.~T., \& Leggett, S.~K.\ 1997, \apjs, 108, 339

\bibitem[Chapman(1930)]{chap30} Chapman, S. A. (1930) Theory of Upper-Atmospheric Ozone, Mem.\ R.\ Met. Soc., 3, 26, pp 103-125.

\bibitem[Cockell(1998)]{cock98}Cockell, C. S. (1998), Biological Journal of the Linnean Society, 63: 449-457

\bibitem[Demarais et al.(2012)]{dema12} Demarais, N.~J., Yang, Z., Martinez, O., et al.\ 2012, \apj, 746, 32

\bibitem[Diffey(1991)]{diffe91}Diffey B.L. Rev Phys Med Biol 1991: 36(3): 299?328.

\bibitem[Farihi(2016)]{fari16} Farihi, J.\ 2016, \nar, 71, 9

\bibitem[Farihi et al.(2013)]{fari13} Farihi, J., G{\"a}nsicke, B.~T., \& Koester, D.\ 2013, Science, 342, 218

\bibitem[Fontaine et al.(2001)]{fbb2001} Fontaine, G., Brassard, P., \& Bergeron, P.\ 2001, \pasp, 113, 409 

\bibitem[Fossati et al.(2012)]{foss12} Fossati, L., Bagnulo, S., Haswell, C.~A., et al.\ 2012, \apjl, 757, L15

\bibitem[Fossati et al.(2015)]{foss15} Fossati, L., Bagnulo, S., Haswell, C.~A., et al.\ 2015, Polarimetry, 305, 325

\bibitem[Fulton et al.(2014)]{fult14} Fulton, B.~J., Tonry, J.~L., Flewelling, H., et al.\ 2014, \apj, 796, 114

\bibitem[Giammichele et al.(2012)]{giam12} Giammichele, N., Bergeron, P., \& Dufour, P.\ 2012, \apjs, 199, 29

\bibitem[Haghighipour \& Kaltenegger(2013)]{hagh13} Haghighipour, N., \& Kaltenegger, L.\ 2013, \apj, 777, 166

\bibitem[Hamers \& Portegies Zwart(2016)]{hame16} Hamers, A.~S., \& Portegies Zwart, S.~F.\ 2016, \mnras, 462, L84

\bibitem[Haqq-Misra et al.(2008)]{haqq08} Haqq-Misra, J.~D., Domagal-Goldman, S.~D., Kasting, P.~J., \& Kasting, J.~F.\ 2008, Astrobiology, 8, 1127

\bibitem[Hougton et al.(2004)]{houg04}Houghton, J.T., Meira Filho, L.G., Bruce, J., Lee, H., Callander, B.A., Haites, E., Harris, N., and Maskell, K. (eds.) (1994) \emph{Climate Change, 1994: Radiative Forcing of Climate Change and an Evaluation of the IPCC IS92 Emission Scenarios}, Cambridge University Press, Cambridge, UK.

\bibitem[Howell et al.(2014)]{howe14} Howell, S.~B., Sobeck, C., Haas, M., et al.\ 2014, \pasp, 126, 398

\bibitem[Jura \& Xu(2010)]{jura10} Jura, M., \& Xu, S.\ 2010, \aj, 140, 1129

\bibitem[Jura \& Young(2014)]{jura14} Jura, M., \& Young, E.~D.\ 2014, Annual Review of Earth and Planetary Sciences, 42, 45

\bibitem[Kaltenegger(2017)]{kalt17} Kaltenegger, L.\ 2017, \araa, 55, 433

\bibitem[Kaltenegger \& Haghighipour(2013)]{kalt13} Kaltenegger, L., \& Haghighipour, N.\ 2013, \apj, 777, 165 

\bibitem[Kaltenegger(2010)]{kalt10} Kaltenegger, L.\ 2010, \apjl, 712, L125

\bibitem[Kaltenegger \& Sasselov(2010)]{kalt10} Kaltenegger, L., \& Sasselov, D.\ 2010, \apj, 708, 1162


\bibitem[Kasting et al.(1993)]{kast93} Kasting, J.~F., Whitmire, D.~P., \& Reynolds, R.~T.\ 1993, \icarus, 101, 108

\bibitem[Kasting \& Ackerman(1986)]{kast86} Kasting, J.~F., \& Ackerman, T.~P.\ 1986, Science, 234, 1383 

\bibitem[Kasting et al.(1985)]{kast85}Kasting, J.~F., Holland, H.~D., Pinto, J.~P. 1985, J. Geophys. Res. 90, 10497?10510

\bibitem[Kasting \& Ackerman(1986)]{kast86} Kasting, J.~F., \& Ackerman, T.~P.\ 1986, Science, 234, 1383

\bibitem[Kepler et al.(2016)]{kpk2016} Kepler, S.~O., Pelisoli, I., Koester, D., et al.\ 2016, \mnras, 455, 3413

\bibitem[Kerwin \& Remmele(2007)]{kerw07}Kerwin BA, Remmele RL, Jr. 2007.  J Pharm Sci. 96(6):1468?1479. 153

\bibitem[Kilic et al.(2009a)]{kili09a} Kilic, M., Kowalski, P.~M., Reach, W.~T., \& von Hippel, T.\ 2009, \apj, 696, 2094

\bibitem[Kilic et al.(2009b)]{kili09b} Kilic, M., Kowalski, P.~M., \& von Hippel, T.\ 2009, \aj, 138, 102

\bibitem[Klein et al.(2011)]{klei11} Klein, B., Jura, M., Koester, D., \& Zuckerman, B.\ 2011, \apj, 741, 64

\bibitem[Koester \& Wilken(2006)]{koes06} Koester, D., \& Wilken, D.\ 2006, \aap, 453, 1051

\bibitem[Koester et al.(2014)]{koes14} Koester, D., G{\"a}nsicke, B.~T., \& Farihi, J.\ 2014, \aap, 566, A34

\bibitem[Kopparapu et al.(2014)]{kopp14} Kopparapu, R.~K., Ramirez, R.~M., SchottelKotte, J., et al.\ 2014, \apjl, 787, L29

\bibitem[Kopparapu et al.(2013)]{kopp13} Kopparapu, R.~K., Ramirez, R., Kasting, J.~F., et al.\ 2013, \apj, 770, 82

\bibitem[Kowalski \& Saumon(2006)]{kowa06} Kowalski, P.~M., \& Saumon, D.\ 2006, \apjl, 651, L137

\bibitem[Kunitomo et al.(2011)]{kuni11} Kunitomo, M., Ikoma, M., Sato, B., Katsuta, Y., \& Ida, S.\ 2011, \apj, 737, 66

\bibitem[Loeb \& Maoz(2013)]{loeb13} Loeb, A., \& Maoz, D.\ 2013, \mnras, 432, 11

\bibitem[Malamud \& Perets(2017b)]{mala17b} Malamud, U., \& Perets, H.~B.\ 2017, \apj, 849, 8 

\bibitem[Malamud \& Perets(2017a)]{mala17a} Malamud, U., \& Perets, H.~B.\ 2017, \apj, 842, 67

\bibitem[Malamud \& Perets(2016)]{mala16} Malamud, U., \& Perets, H.~B.\ 2016, \apj, 832, 160

\bibitem[Matasunaga et al.(1991)]{mats91}Matsunaga, T., Heida, K., and Nikaido, O. (1991), Photochem. Photobiol., 54, 403-410. 19.

\bibitem[McCree(1971)]{mccr71}McCree K. J., 1971, Agricultural Meteorology, 9, 191

\bibitem[Mustill \& Villaver(2012)]{must12} Mustill, A.~J., \& Villaver, E.\ 2012, \apj, 761, 121

\bibitem[Parsons et al.(2017)]{pgm2017} Parsons, S.~G., G{\"a}nsicke, B.~T., Marsh, T.~R., et al.\ 2017, \mnras, 470, 4473

\bibitem[Pavlov \& Kasting(2002)]{pavl02} Pavlov, A.~A., \& Kasting, J.~F.\ 2002, Astrobiology, 2, 27

\bibitem[Pavlov et al.(2000)]{pavl00} Pavlov, A.~A., Kasting, J.~F., Brown, L.~L., Rages, K.~A., \& Freedman, R.\ 2000, \jgr, 105, 11981

\bibitem[Ramirez \& Kaltenegger(2017)]{rami17} Ramirez, R.~M., \& Kaltenegger, L.\ 2017, \apjl, 837, L4

\bibitem[Ramirez \& Kaltenegger(2016)]{rami16} Ramirez, R.~M., \& Kaltenegger, L.\ 2016, \apj, 823, 6

\bibitem[Ramirez \& Kaltenegger(2014)]{rami14} Ramirez, R.~M., \& Kaltenegger, L.\ 2014, \apjl, 797, L25

\bibitem[Rugheimer \& Kaltenegger(2018)]{rugh18} Rugheimer, S., \& Kaltenegger, L.\ 2018, \apj, 854, 19

\bibitem[Rugheimer et al.(2015)]{rugh15} Rugheimer, S., Kaltenegger, L., Segura, A., Linsky, J., \& Mohanty, S.\ 2015, \apj, 809, 57

\bibitem[Rugheimer et al.(2015b)]{rugh15b} Rugheimer, S., Segura, A., Kaltenegger, L., \& Sasselov, D.\ 2015, \apj, 806, 137

\bibitem[Rugheimer et al.(2013)]{rugh13} Rugheimer, S., Kaltenegger, L., Zsom, A., Segura, A., \& Sasselov, D.\ 2013, Astrobiology, 13, 251

\bibitem[Saumon et al.(2014)]{saum14} Saumon, D., Holberg, J.~B., \& Kowalski, P.~M.\ 2014, \apj, 790, 50

\bibitem[Segura et al.(2007)]{segu07} Segura, A., Meadows, V.~S., Kasting, J.~F., Crisp, D., \& Cohen, M.\ 2007, \aap, 472, 665

\bibitem[Segura et al.(2005)]{segu05} Segura, A., Kasting, J.~F., Meadows, V., et al.\ 2005, Astrobiology, 5, 706

\bibitem[Segura et al.(2003)]{segu03} Segura, A., Krelove, K., Kasting, J.~F., et al.\ 2003, Astrobiology, 3, 689

\bibitem[Tevini(1993)]{tevi93}Tevini M (1993), Lewis, Boca Raton, pp 125?154.

\bibitem[Toon et al.(1989)]{toon89} Toon, O.~B., McKay, C.~P., Ackerman, T.~P., \& Santhanam, K.\ 1989, \jgr, 94, 16287


\bibitem[Van Eylen et al.(2017)]{vane17} Van Eylen, V., Agentoft, C., Lundkvist, M.~S., et al.\ 2017, arXiv:1710.05398

\bibitem[van Sluijs \& Van Eylen(2017)]{slui17} van Sluijs, L., \& Van Eylen, V.\ 2017, arXiv:1711.09691 

\bibitem[Vanderburg et al.(2015)]{vand15} Vanderburg, A., Johnson, J.~A., Rappaport, S., et al.\ 2015, \nat, 526, 546

\bibitem[Veras \& G{\"a}nsicke(2015)]{vera15} Veras, D., \& G{\"a}nsicke, B.~T.\ 2015, \mnras, 447, 1049

\bibitem[Veras(2016)]{vera16} Veras, D.\ 2016, Royal Society Open Science, 3, 150571

\bibitem[Villaver et al.(2014)]{villa14} Villaver, E., Livio, M., Mustill, A.~J., \& Siess, L.\ 2014, \apj, 794, 3

\bibitem[Villaver(2012)]{villa12} Villaver, E.\ 2012, IAU Symposium, 283, 219

\bibitem[Villaver(2011)]{villa11} Villaver, E.\ 2011, American Institute of Physics Conference Series, 1331, 21

\bibitem[Voet et al.(1963)]{voet63} Voet D, Gratzer W.B., Cox R. A.,  and Doty P..,  Biopolymers, Vol. 1, 1963 , pp. 193-208.

\bibitem[Wolszczan \& Frail(1992)]{wols92} Wolszczan, A., \& Frail, D.~A.\ 1992, \nat, 355, 145 

\bibitem[Xu et al.(2015)]{xu15} Xu, S., Ertel, S., Wahhaj, Z., et al.\ 2015, \aap, 579, L8


\end{thebibliography}
\end{document}